
\expandafter\ifx\csname phyzzx\endcsname\relax
 \message{It is better to use PHYZZX format than to
          \string\input\space PHYZZX}\else
 \wlog{PHYZZX macros are already loaded and are not
          \string\input\space again}%
 \endinput \fi
\catcode`\@=11 
\let\rel@x=\relax
\let\n@expand=\relax
\def\pr@tect{\let\n@expand=\noexpand}
\let\protect=\pr@tect
\let\gl@bal=\global
%
\newfam\cpfam
\newdimen\b@gheight             \b@gheight=12pt
\newcount\f@ntkey               \f@ntkey=0
\def\f@m{\afterassignment\samef@nt\f@ntkey=}
\def\samef@nt{\fam=\f@ntkey \the\textfont\f@ntkey\rel@x}
\def\setstr@t{\setbox\strutbox=\hbox{\vrule height 0.85\b@gheight
                                depth 0.35\b@gheight width\z@ }}

\font\fourteenrm  =cmr10 scaled\magstep2
\font\twelverm    =cmr10 scaled\magstep1
\font\ninerm      =cmr9
\font\sixrm       =cmr6

\font\fourteenbf  =cmbx10 scaled\magstep3
\font\twelvebf    =cmbx10 scaled\magstep1
\font\ninebf      =cmbx9
\font\sixbf       =cmbx6
\font\seventeeni  =cmmi10 scaled\magstep3    \skewchar\seventeeni='177
\font\fourteeni   =cmmi10 scaled\magstep2     \skewchar\fourteeni='177
\font\twelvei     =cmmi10 scaled\magstep1       \skewchar\twelvei='177
\font\ninei       =cmmi9                          \skewchar\ninei='177
\font\sixi        =cmmi6                           \skewchar\sixi='177
\font\seventeensy =cmsy10 scaled\magstep3    \skewchar\seventeensy='60
\font\fourteensy  =cmsy10 scaled\magstep2     \skewchar\fourteensy='60
\font\twelvesy    =cmsy10 scaled\magstep1       \skewchar\twelvesy='60
\font\ninesy      =cmsy9                          \skewchar\ninesy='60
\font\sixsy       =cmsy6                           \skewchar\sixsy='60

\font\fourteenex  =cmex10 scaled\magstep2
\font\twelveex    =cmex10 scaled\magstep1

\font\fourteensl  =cmsl10 scaled\magstep2
\font\twelvesl    =cmsl10 scaled\magstep1
\font\ninesl      =cmsl9

\font\fourteenit  =cmti10 scaled\magstep2
\font\twelveit    =cmti10 scaled\magstep1
\font\nineit      =cmti9
\font\fourteentt  =cmtt10 scaled\magstep2
\font\twelvett    =cmtt10 scaled\magstep1
\font\fourteencp  =cmcsc10 scaled\magstep2
\font\twelvecp    =cmcsc10 scaled\magstep1
\font\tencp       =cmcsc10
%
%
\def\fourteenf@nts{\relax
    \textfont0=\fourteenrm          \scriptfont0=\tenrm
      \scriptscriptfont0=\sevenrm
    \textfont1=\fourteeni           \scriptfont1=\teni
      \scriptscriptfont1=\seveni
    \textfont2=\fourteensy          \scriptfont2=\tensy
      \scriptscriptfont2=\sevensy
    \textfont3=\fourteenex          \scriptfont3=\twelveex
      \scriptscriptfont3=\tenex
    \textfont\itfam=\fourteenit     \scriptfont\itfam=\tenit
    \textfont\slfam=\fourteensl     \scriptfont\slfam=\tensl
    \textfont\bffam=\fourteenbf     \scriptfont\bffam=\tenbf
      \scriptscriptfont\bffam=\sevenbf
    \textfont\ttfam=\fourteentt
    \textfont\cpfam=\fourteencp }
\def\twelvef@nts{\relax
    \textfont0=\twelverm          \scriptfont0=\ninerm
      \scriptscriptfont0=\sixrm
    \textfont1=\twelvei           \scriptfont1=\ninei
      \scriptscriptfont1=\sixi
    \textfont2=\twelvesy           \scriptfont2=\ninesy
      \scriptscriptfont2=\sixsy
    \textfont3=\twelveex          \scriptfont3=\tenex
      \scriptscriptfont3=\tenex
    \textfont\itfam=\twelveit     \scriptfont\itfam=\nineit
    \textfont\slfam=\twelvesl     \scriptfont\slfam=\ninesl
    \textfont\bffam=\twelvebf     \scriptfont\bffam=\ninebf
      \scriptscriptfont\bffam=\sixbf
    \textfont\ttfam=\twelvett
    \textfont\cpfam=\twelvecp }
\def\tenf@nts{\relax
    \textfont0=\tenrm          \scriptfont0=\sevenrm
      \scriptscriptfont0=\fiverm
    \textfont1=\teni           \scriptfont1=\seveni
      \scriptscriptfont1=\fivei
    \textfont2=\tensy          \scriptfont2=\sevensy
      \scriptscriptfont2=\fivesy
    \textfont3=\tenex          \scriptfont3=\tenex
      \scriptscriptfont3=\tenex
    \textfont\itfam=\tenit     \scriptfont\itfam=\seveni  
    \textfont\slfam=\tensl     \scriptfont\slfam=\sevenrm 
    \textfont\bffam=\tenbf     \scriptfont\bffam=\sevenbf
      \scriptscriptfont\bffam=\fivebf
    \textfont\ttfam=\tentt
    \textfont\cpfam=\tencp }
%
%

%
\def\rm{\n@expand\f@m0 }
\def\mit{\n@expand\f@m1 }         
\def\cal{\n@expand\f@m2 }
\def\it{\n@expand\f@m\itfam}
\def\sl{\n@expand\f@m\slfam}
\def\bf{\n@expand\f@m\bffam}
\def\tt{\n@expand\f@m\ttfam}
\def\caps{\n@expand\f@m\cpfam}    
\def\em@{\rel@x\ifnum\f@ntkey=0 \it \else
        \ifnum\f@ntkey=\bffam \it \else \rm \fi \fi }
\def\em{\n@expand\em@}
\def\fourteenpoint{\fourteenf@nts \samef@nt \b@gheight=14pt \setstr@t }
\def\twelvepoint{\twelvef@nts \samef@nt \b@gheight=12pt \setstr@t }
\def\tenpoint{\tenf@nts \samef@nt \b@gheight=10pt \setstr@t }
\normalbaselineskip = 20pt plus 0.2pt minus 0.1pt
\normallineskip = 1.5pt plus 0.1pt minus 0.1pt
\normallineskiplimit = 1.5pt
\newskip\normaldisplayskip
\normaldisplayskip = 20pt plus 5pt minus 10pt
\newskip\normaldispshortskip
\normaldispshortskip = 6pt plus 5pt
\newskip\normalparskip
\normalparskip = 6pt plus 2pt minus 1pt
\newskip\skipregister
\skipregister = 5pt plus 2pt minus 1.5pt
\newif\ifsingl@
\newif\ifdoubl@
\newif\iftwelv@  \twelv@true
\def\singlespace{\singl@true\doubl@false\spaces@t}
\def\doublespace{\singl@false\doubl@true\spaces@t}
\def\normalspace{\singl@false\doubl@false\spaces@t}
\def\Tenpoint{\tenpoint\twelv@false\spaces@t}
\def\Twelvepoint{\twelvepoint\twelv@true\spaces@t}
\def\spaces@t{\rel@x
      \iftwelv@ \ifsingl@\subspaces@t3:4;\else\subspaces@t1:1;\fi
       \else \ifsingl@\subspaces@t3:5;\else\subspaces@t4:5;\fi \fi
      \ifdoubl@ \multiply\baselineskip by 5
         \divide\baselineskip by 4 \fi }
\def\subspaces@t#1:#2;{
      \baselineskip = \normalbaselineskip
      \multiply\baselineskip by #1 \divide\baselineskip by #2
      \lineskip = \normallineskip
      \multiply\lineskip by #1 \divide\lineskip by #2
      \lineskiplimit = \normallineskiplimit
      \multiply\lineskiplimit by #1 \divide\lineskiplimit by #2
      \parskip = \normalparskip
      \multiply\parskip by #1 \divide\parskip by #2
      \abovedisplayskip = \normaldisplayskip
      \multiply\abovedisplayskip by #1 \divide\abovedisplayskip by #2
      \belowdisplayskip = \abovedisplayskip
      \abovedisplayshortskip = \normaldispshortskip
      \multiply\abovedisplayshortskip by #1
        \divide\abovedisplayshortskip by #2
      \belowdisplayshortskip = \abovedisplayshortskip
      \advance\belowdisplayshortskip by \belowdisplayskip
      \divide\belowdisplayshortskip by 2
      \smallskipamount = \skipregister
      \multiply\smallskipamount by #1 \divide\smallskipamount by #2
      \medskipamount = \smallskipamount \multiply\medskipamount by 2
      \bigskipamount = \smallskipamount \multiply\bigskipamount by 4 }
\def\normalbaselines{ \baselineskip=\normalbaselineskip
   \lineskip=\normallineskip \lineskiplimit=\normallineskip
   \iftwelv@\else \multiply\baselineskip by 4 \divide\baselineskip by 5
     \multiply\lineskiplimit by 4 \divide\lineskiplimit by 5
     \multiply\lineskip by 4 \divide\lineskip by 5 \fi }
\Twelvepoint  
\interlinepenalty=50
\interfootnotelinepenalty=5000
\predisplaypenalty=9000
\postdisplaypenalty=500
\hfuzz=1pt
\vfuzz=0.2pt
\newdimen\HOFFSET  \HOFFSET=0pt
\newdimen\VOFFSET  \VOFFSET=0pt
\newdimen\HSWING   \HSWING=0pt
\dimen\footins=8in
%
%
%
\newskip\pagebottomfiller
\pagebottomfiller=\z@ plus \z@ minus \z@
\def\pagecontents{
   \ifvoid\topins\else\unvbox\topins\vskip\skip\topins\fi
   \dimen@ = \dp255 \unvbox255
   \vskip\pagebottomfiller
   \ifvoid\footins\else\vskip\skip\footins\footrule\unvbox\footins\fi
   \ifr@ggedbottom \kern-\dimen@ \vfil \fi }
\def\makeheadline{\vbox to 0pt{ \skip@=\topskip
      \advance\skip@ by -12pt \advance\skip@ by -2\normalbaselineskip
      \vskip\skip@ \line{\vbox to 12pt{}\the\headline} \vss
      }\nointerlineskip}
\def\makefootline{\baselineskip = 1.5\normalbaselineskip
                 \line{\the\footline}}
\newif\iffrontpage
\newif\ifp@genum
\def\nopagenumbers{\p@genumfalse}
\def\pagenumbers{\p@genumtrue}
\pagenumbers
\newtoks\paperheadline
\newtoks\paperfootline
\newtoks\letterheadline
\newtoks\letterfootline
\newtoks\letterinfo
\newtoks\date
\paperheadline={\hfil}
\paperfootline={\hss\iffrontpage\else\ifp@genum\tenrm\folio\hss\fi\fi}
\letterheadline{\iffrontpage \hfil \else
    \rm \ifp@genum page~~\folio\fi \hfil\the\date \fi}
\letterfootline={\iffrontpage\the\letterinfo\else\hfil\fi}
\letterinfo={\hfil}
\def\monthname{\rel@x\ifcase\month 0/\or January\or February\or
   March\or April\or May\or June\or July\or August\or September\or
   October\or November\or December\else\number\month/\fi}
\def\today{\monthname~\number\day, \number\year}
\date={\today}
\headline=\paperheadline 
\footline=\paperfootline 
\countdef\pageno=1      \countdef\pagen@=0
\countdef\pagenumber=1  \pagenumber=1
\def\advancepageno{\gl@bal\advance\pagen@ by 1
   \ifnum\pagenumber<0 \gl@bal\advance\pagenumber by -1
    \else\gl@bal\advance\pagenumber by 1 \fi
    \gl@bal\frontpagefalse  \swing@ }
\def\folio{\ifnum\pagenumber<0 \romannumeral-\pagenumber
           \else \number\pagenumber \fi }
\def\swing@{\ifodd\pagenumber \gl@bal\advance\hoffset by -\HSWING
             \else \gl@bal\advance\hoffset by \HSWING \fi }
\def\footrule{\dimen@=\prevdepth\nointerlineskip
   \vbox to 0pt{\vskip -0.25\baselineskip \hrule width 0.35\hsize \vss}
   \prevdepth=\dimen@ }
\let\footnotespecial=\rel@x
\newdimen\footindent
\footindent=24pt
\def\Textindent#1{\noindent\llap{#1\enspace}\ignorespaces}
\def\Vfootnote#1{\insert\footins\bgroup
   \interlinepenalty=\interfootnotelinepenalty \floatingpenalty=20000
   \singl@true\doubl@false\Tenpoint
   \splittopskip=\ht\strutbox \boxmaxdepth=\dp\strutbox
   \leftskip=\footindent \rightskip=\z@skip
   \parindent=0.5\footindent \parfillskip=0pt plus 1fil
   \spaceskip=\z@skip \xspaceskip=\z@skip \footnotespecial
   \Textindent{#1}\footstrut\futurelet\next\fo@t}

\def\vfootnote#1{\Vfootnote{${#1}$}}
\def\footnote#1{\attach{#1}\vfootnote{#1}}

\let\footsymbol=\star
\newcount\lastf@@t           \lastf@@t=-1
\newcount\footsymbolcount    \footsymbolcount=0
\newif\ifPhysRev
\def\bumpfootsymbolcount{\rel@x
   \iffrontpage \bumpfootsymbolpos \else \advance\lastf@@t by 1
     \ifPhysRev \bumpfootsymbolneg \else \bumpfootsymbolpos \fi \fi
   \gl@bal\lastf@@t=\pagen@ }
\def\bumpfootsymbolpos{\ifnum\footsymbolcount <0
                            \gl@bal\footsymbolcount =0 \fi
    \ifnum\lastf@@t<\pagen@ \gl@bal\footsymbolcount=0
     \else \gl@bal\advance\footsymbolcount by 1 \fi }
\def\bumpfootsymbolneg{\ifnum\footsymbolcount >0
             \gl@bal\footsymbolcount =0 \fi
         \gl@bal\advance\footsymbolcount by -1 }
\def\fd@f#1 {\xdef\footsymbol{\mathchar"#1 }}
\def\generatefootsymbol{\ifcase\footsymbolcount \fd@f 13F \or \fd@f 279
        \or \fd@f 27A \or \fd@f 278 \or \fd@f 27B \else
        \ifnum\footsymbolcount <0 \fd@f{023 \number-\footsymbolcount }
         \else \fd@f 203 {\loop \ifnum\footsymbolcount >5
                \fd@f{203 \footsymbol } \advance\footsymbolcount by -1
                \repeat }\fi \fi }

\def\nonfrenchspacing{\sfcode`\.=3001 \sfcode`\!=3000 \sfcode`\?=3000
        \sfcode`\:=2000 \sfcode`\;=1500 \sfcode`\,=1251 }
\nonfrenchspacing
\newdimen\d@twidth
{\setbox0=\hbox{s.} \gl@bal\d@twidth=\wd0 \setbox0=\hbox{s}
        \gl@bal\advance\d@twidth by -\wd0 }
\def\removehglue{\loop \unskip \ifdim\lastskip >\z@ \repeat }
\def\roll@ver#1{\removehglue \nobreak \count255 =\spacefactor \dimen@=\z@
        \ifnum\count255 =3001 \dimen@=\d@twidth \fi
        \ifnum\count255 =1251 \dimen@=\d@twidth \fi
    \iftwelv@ \kern-\dimen@ \else \kern-0.83\dimen@ \fi
   #1\spacefactor=\count255 }
\def\step@ver#1{\rel@x \ifmmode #1\else \ifhmode
        \roll@ver{${}#1$}\else {\setbox0=\hbox{${}#1$}}\fi\fi }
\def\attach#1{\step@ver{\strut^{\mkern 2mu #1} }}
%
%
%
\newcount\chapternumber      \chapternumber=0
\newcount\sectionnumber      \sectionnumber=0
\newcount\equanumber         \equanumber=0
\let\chapterlabel=\rel@x
\let\sectionlabel=\rel@x
\newtoks\chapterstyle        \chapterstyle={\Number}
\newtoks\sectionstyle        \sectionstyle={\Number}
\newskip\chapterskip         \chapterskip=\bigskipamount
\newskip\sectionskip         \sectionskip=\medskipamount
\newskip\headskip            \headskip=8pt plus 3pt minus 3pt
\newdimen\chapterminspace    \chapterminspace=15pc
\newdimen\sectionminspace    \sectionminspace=10pc
\newdimen\referenceminspace  \referenceminspace=20pc
\newif\ifcn@                 \cn@true
\newif\ifcn@@                \cn@@false
\def\numberedchapters{\cn@true}
\def\unnumberedchapters{\cn@false\sequentialequations}
\def\chapterreset{\gl@bal\advance\chapternumber by 1
   \ifnum\equanumber<0 \else\gl@bal\equanumber=0\fi
   \sectionnumber=0 \let\sectionlabel=\rel@x
   \ifcn@ \gl@bal\cn@@true {\pr@tect
       \xdef\chapterlabel{\the\chapterstyle{\the\chapternumber}}}%
    \else \gl@bal\cn@@false \gdef\chapterlabel{\rel@x}\fi }
\def\@alpha#1{\count255='140 \advance\count255 by #1\char\count255}
 \def\alphabetic{\n@expand\@alpha}
\def\@Alpha#1{\count255='100 \advance\count255 by #1\char\count255}
 \def\Alphabetic{\n@expand\@Alpha}
\def\@Roman#1{\uppercase\expandafter{\romannumeral #1}}
 \def\Roman{\n@expand\@Roman}
\def\@roman#1{\romannumeral #1}    \def\roman{\n@expand\@roman}
\def\@number#1{\number #1}         \def\Number{\n@expand\@number}
\def\BLANK#1{\rel@x}               
\def\titleparagraphs{\interlinepenalty=9999
     \leftskip=0.03\hsize plus 0.22\hsize minus 0.03\hsize
     \rightskip=\leftskip \parfillskip=0pt
     \hyphenpenalty=9000 \exhyphenpenalty=9000
     \tolerance=9999 \pretolerance=9000
     \spaceskip=0.333em \xspaceskip=0.5em }
\def\titlestyle#1{\par\begingroup \titleparagraphs
     \iftwelv@\fourteenpoint\else\twelvepoint\fi
   \noindent #1\par\endgroup }
\def\spacecheck#1{\dimen@=\pagegoal\advance\dimen@ by -\pagetotal
   \ifdim\dimen@<#1 \ifdim\dimen@>0pt \vfil\break \fi\fi}
\def\chapter#1{\par \penalty-300 \vskip\chapterskip
   \spacecheck\chapterminspace
   \chapterreset \titlestyle{\ifcn@@\chapterlabel.~\fi #1}
   \nobreak\vskip\headskip \penalty 30000
   {\pr@tect\wlog{\string\chapter\space \chapterlabel}} }

%


\def\section#1{\par \ifnum\lastpenalty=30000\else
   \penalty-200\vskip\sectionskip \spacecheck\sectionminspace\fi
   \gl@bal\advance\sectionnumber by 1
   {\pr@tect
   \ifcn@@\expandafter\toks@\expandafter{\chapterlabel.}\else\toks@={}\fi
   \xdef\sectionlabel{\the\toks@\the\sectionstyle{\the\sectionnumber}}%
   \wlog{\string\section\space \sectionlabel}}%
   \noindent {\bf\enspace\sectionlabel.~~#1}\par
   \nobreak\vskip\headskip \penalty 30000 }

\def\subsection#1{\par
   \ifnum\the\lastpenalty=30000\else \penalty-100\smallskip \fi
   \noindent\undertext{#1}\enspace \vadjust{\penalty5000}}

\def\undertext#1{\vtop{\hbox{#1}\kern 1pt \hrule}}

\def\ack{\subsection{Acknowledgements:}}
\def\APPENDIX#1#2{\par\penalty-300\vskip\chapterskip
   \spacecheck\chapterminspace \chapterreset \xdef\chapterlabel{#1}
   \titlestyle{APPENDIX #2} \nobreak\vskip\headskip \penalty 30000
   \wlog{\string\Appendix~\chapterlabel} }
\def\Appendix#1{\APPENDIX{#1}{#1}}
\def\appendix{\APPENDIX{A}{}}
%
%
%
%
%
\def\eqname#1{\rel@x {\pr@tect
  \ifnum\equanumber<0 \xdef#1{{(\number-\equanumber)}}%
     \gl@bal\advance\equanumber by -1
  \else \gl@bal\advance\equanumber by 1
   \ifcn@@ \toks@=\expandafter{\chapterlabel.}\else\toks@={}\fi
   \xdef#1{{(\the\toks@\number\equanumber)}}\fi #1}}

\def\eq{\eqname\?}
\def\eqn{\eqno\eqname}

\def\eqinsert#1{\noalign{\dimen@=\prevdepth \nointerlineskip
   \setbox0=\hbox to\displaywidth{\hfil #1}
   \vbox to 0pt{\kern 0.5\baselineskip\hbox{$\!\box0\!$}\vss}
   \prevdepth=\dimen@}}
%

%
%
\def\GENITEM#1;#2{\par \hangafter=0 \hangindent=#1
    \Textindent{$ #2 $}\ignorespaces}
\outer\def\newitem#1=#2;{\gdef#1{\GENITEM #2;}}

\newdimen\itemsize                \itemsize=30pt
\newitem\item=1\itemsize;
\newitem\sitem=1.75\itemsize;     
\newitem\ssitem=2.5\itemsize;     
\outer\def\newlist#1=#2&#3&#4;{\toks0={#2}\toks1={#3}%
   \count255=\escapechar \escapechar=-1
   \alloc@0\list\countdef\insc@unt\listcount     \listcount=0
   \edef#1{\par
      \countdef\listcount=\the\allocationnumber
      \advance\listcount by 1
      \hangafter=0 \hangindent=#4
      \Textindent{\the\toks0{\listcount}\the\toks1}}
   \expandafter\expandafter\expandafter
    \edef\c@t#1{begin}{\par
      \countdef\listcount=\the\allocationnumber \listcount=1
      \hangafter=0 \hangindent=#4
      \Textindent{\the\toks0{\listcount}\the\toks1}}
   \expandafter\expandafter\expandafter
    \edef\c@t#1{con}{\par \hangafter=0 \hangindent=#4 \noindent}
   \escapechar=\count255}
\def\c@t#1#2{\csname\string#1#2\endcsname}
\newlist\point=\Number&.&1.0\itemsize;
\newlist\subpoint=(\alphabetic&)&1.75\itemsize;
\newlist\subsubpoint=(\roman&)&2.5\itemsize;
%

%
%
%
%
\newcount\referencecount     \referencecount=0
\newcount\lastrefsbegincount \lastrefsbegincount=0
\newif\ifreferenceopen       \newwrite\referencewrite
\newdimen\refindent          \refindent=30pt
\def\normalrefmark#1{\attach{\scriptscriptstyle [ #1 ] }}
\let\PRrefmark=\attach
\def\NPrefmark#1{\step@ver{{\;[#1]}}}
\def\refmark#1{\rel@x\ifPhysRev\PRrefmark{#1}\else\normalrefmark{#1}\fi}
\def\refend@{\refmark{\number\referencecount}}
\def\refend{\refend@{}\space }
\def\refsend{\refmark{\count255=\referencecount
   \advance\count255 by-\lastrefsbegincount
   \ifcase\count255 \number\referencecount
   \or \number\lastrefsbegincount,\number\referencecount
   \else \number\lastrefsbegincount-\number\referencecount \fi}\space }
\def\REFNUM#1{\rel@x \gl@bal\advance\referencecount by 1
    \xdef#1{\the\referencecount }}
\def\Refnum#1{\REFNUM #1\refend@ } \let\refnum=\Refnum
\def\REF#1{\REFNUM #1\R@FWRITE\ignorespaces}
\def\Ref#1{\Refnum #1\REFWRITE }
\def\ref{\Ref\?}
\def\REFS#1{\REFNUM #1\gl@bal\lastrefsbegincount=\referencecount
    \REFWRITE }

       \let\REFSCON=\REF
\def\r@fitem#1{\par \hangafter=0 \hangindent=\refindent \Textindent{#1}}
\def\refitem#1{\r@fitem{#1.}}
\def\NPrefitem#1{\r@fitem{[#1]}}
\def\NPrefs{\let\refmark=\NPrefmark \let\refitem=NPrefitem}
\def\REFWRITE{\R@FWRITE\rel@x }
\def\R@FWRITE#1{\ifreferenceopen \else \gl@bal\referenceopentrue
     \immediate\openout\referencewrite=\jobname.refs
     \toks@={\begingroup \refoutspecials \catcode`\^^M=10 }%
     \immediate\write\referencewrite{\the\toks@}\fi
    \immediate\write\referencewrite{\noexpand\refitem %
                                    {\the\referencecount}}%
    \p@rse@ndwrite \referencewrite #1}
\begingroup
 \catcode`\^^M=\active \let^^M=\relax %
 \gdef\p@rse@ndwrite#1#2{\begingroup \catcode`\^^M=12 \newlinechar=`\^^M%
         \chardef\rw@write=#1\sc@nlines#2}%
 \gdef\sc@nlines#1#2{\sc@n@line \g@rbage #2^^M\endsc@n \endgroup #1}%
 \gdef\sc@n@line#1^^M{\expandafter\toks@\expandafter{\deg@rbage #1}%
         \immediate\write\rw@write{\the\toks@}%
         \futurelet\n@xt \sc@ntest }%
\endgroup
\def\sc@ntest{\ifx\n@xt\endsc@n \let\n@xt=\rel@x
       \else \let\n@xt=\sc@n@notherline \fi \n@xt }
\def\sc@n@notherline{\sc@n@line \g@rbage }
\def\deg@rbage#1{}
\let\g@rbage=\relax    \let\endsc@n=\relax
\def\refout{\par\penalty-400\vskip\chapterskip
   \spacecheck\referenceminspace
   \ifreferenceopen \Closeout\referencewrite \referenceopenfalse \fi
   \line{\fourteenrm\hfil REFERENCES\hfil}\vskip\headskip
   \input \jobname.refs
   }

\def\myrefout{\par\penalty-400
   \spacecheck\referenceminspace
    \ifreferenceopen \Closeout\referencewrite \referenceopenfalse \fi
   \input \jobname.refs
   }

\def\refoutspecials{\sfcode`\.=1000 \interlinepenalty=1000
         \rightskip=\z@ plus 1em minus \z@ }
\def\Closeout#1{\toks0={\par\endgroup}\immediate\write#1{\the\toks0}%
   \immediate\closeout#1}
%
%
\newcount\figurecount     \figurecount=0
\newcount\tablecount      \tablecount=0
\newif\iffigureopen       \newwrite\figurewrite
\newif\iftableopen        \newwrite\tablewrite
\def\FIGNUM#1{\rel@x \gl@bal\advance\figurecount by 1
    \xdef#1{\the\figurecount}}
\def\FIGURE#1{\FIGNUM #1\F@GWRITE\ignorespaces }
\let\FIG=\FIGURE

\def\figitem#1{\r@fitem{#1)}}
\def\FIGWRITE{\F@GWRITE\rel@x }
\def\TABNUM#1{\rel@x \gl@bal\advance\tablecount by 1
    \xdef#1{\the\tablecount}}
\def\TABLE#1{\TABNUM #1\T@BWRITE\ignorespaces }

\def\tabitem#1{\r@fitem{#1:}}
\def\TABWRITE{\T@BWRITE\rel@x }
\def\F@GWRITE#1{\iffigureopen \else \gl@bal\figureopentrue
     \immediate\openout\figurewrite=\jobname.figs
     \toks@={\begingroup \catcode`\^^M=10 }%
     \immediate\write\figurewrite{\the\toks@}\fi
    \immediate\write\figurewrite{\noexpand\figitem %
                                 {\the\figurecount}}%
    \p@rse@ndwrite \figurewrite #1}
\def\T@BWRITE#1{\iftableopen \else \gl@bal\tableopentrue
     \immediate\openout\tablewrite=\jobname.tabs
     \toks@={\begingroup \catcode`\^^M=10 }%
     \immediate\write\tablewrite{\the\toks@}\fi
    \immediate\write\tablewrite{\noexpand\tabitem %
                                 {\the\tablecount}}%
    \p@rse@ndwrite \tablewrite #1}
\def\figout{\par\penalty-400
   \vskip\chapterskip\spacecheck\referenceminspace
   \iffigureopen \Closeout\figurewrite \figureopenfalse \fi
   \line{\fourteenrm\hfil FIGURE CAPTIONS\hfil}\vskip\headskip
   \input \jobname.figs
   }
\def\tabout{\par\penalty-400
   \vskip\chapterskip\spacecheck\referenceminspace
   \iftableopen \Closeout\tablewrite \tableopenfalse \fi
   \line{\fourteenrm\hfil TABLE CAPTIONS\hfil}\vskip\headskip
   \input \jobname.tabs
   }
%
%
%
\newbox\picturebox
\def\p@cht{\ht\picturebox }
\def\p@cwd{\wd\picturebox }
\def\p@cdp{\dp\picturebox }
\newdimen\xshift
\newdimen\yshift
\newdimen\captionwidth
\newskip\captionskip
\captionskip=15pt plus 5pt minus 3pt
\def\fullwidth{\captionwidth=\hsize }
\newtoks\Caption
\newif\ifcaptioned
\newif\ifselfcaptioned
\def\caption{\captionedtrue \Caption }
\newcount\linesabove
\newif\iffileexists
\newtoks\picfilename
\def\fil@#1 {\fileexiststrue \picfilename={#1}}
\def\file#1{\if=#1\let\n@xt=\fil@ \else \def\n@xt{\fil@ #1}\fi \n@xt }
\def\pl@t{\begingroup \pr@tect
    \setbox\picturebox=\hbox{}\fileexistsfalse
    \let\height=\p@cht \let\width=\p@cwd \let\depth=\p@cdp
    \xshift=\z@ \yshift=\z@ \captionwidth=\z@
    \Caption={}\captionedfalse
    \linesabove =0 \picturedefault }
\def\plot{\pl@t \selfcaptionedfalse }
\def\Picture#1{\gl@bal\advance\figurecount by 1
    \xdef#1{\the\figurecount}\pl@t \selfcaptionedtrue }

\def\s@vepicture{\iffileexists \parsefilename \redopicturebox \fi
   \ifdim\captionwidth>\z@ \else \captionwidth=\p@cwd \fi
   \xdef\lastpicture{\iffileexists
        \setbox0=\hbox{\raise\the\yshift \vbox{%
              \moveright\the\xshift\hbox{\picturedefinition}}}%
        \else \setbox0=\hbox{}\fi
         \ht0=\the\p@cht \wd0=\the\p@cwd \dp0=\the\p@cdp
         \vbox{\hsize=\the\captionwidth \line{\hss\box0 \hss }%
              \ifcaptioned \vskip\the\captionskip \noexpand\Tenpoint
                \ifselfcaptioned Figure~\the\figurecount.\enspace \fi
                \the\Caption \fi }}%
    \endgroup }
\let\endpicture=\s@vepicture
\def\savepicture#1{\s@vepicture \global\let#1=\lastpicture }
\def\displaypicture{\fullwidth \s@vepicture $$\lastpicture $${}}
\def\toppicture{\fullwidth \s@vepicture \topinsert
    \lastpicture \medskip \endinsert }
\def\midpicture{\fullwidth \s@vepicture \midinsert
    \lastpicture \endinsert }
%
%
\def\leftpicture{\pres@tpicture
    \dimen@i=\hsize \advance\dimen@i by -\dimen@ii
    \setbox\picturebox=\hbox to \hsize {\box0 \hss }%
    \wr@paround }
\def\rightpicture{\pres@tpicture
    \dimen@i=\z@
    \setbox\picturebox=\hbox to \hsize {\hss \box0 }%
    \wr@paround }
\def\pres@tpicture{\gl@bal\linesabove=\linesabove
    \s@vepicture \setbox\picturebox=\vbox{
         \kern \linesabove\baselineskip \kern 0.3\baselineskip
         \lastpicture \kern 0.3\baselineskip }%
    \dimen@=\p@cht \dimen@i=\dimen@
    \advance\dimen@i by \pagetotal
    \par \ifdim\dimen@i>\pagegoal \vfil\break \fi
    \dimen@ii=\hsize
    \advance\dimen@ii by -\parindent \advance\dimen@ii by -\p@cwd
    \setbox0=\vbox to\z@{\kern-\baselineskip \unvbox\picturebox \vss }}
\def\wr@paround{\Caption={}\count255=1
    \loop \ifnum \linesabove >0
         \advance\linesabove by -1 \advance\count255 by 1
         \advance\dimen@ by -\baselineskip
         \expandafter\Caption \expandafter{\the\Caption \z@ \hsize }%
      \repeat
    \loop \ifdim \dimen@ >\z@
         \advance\count255 by 1 \advance\dimen@ by -\baselineskip
         \expandafter\Caption \expandafter{%
             \the\Caption \dimen@i \dimen@ii }%
      \repeat
    \edef\n@xt{\parshape=\the\count255 \the\Caption \z@ \hsize }%
    \par\noindent \n@xt \strut \vadjust{\box\picturebox }}
\let\picturedefault=\relax
\let\parsefilename=\relax
\def\redopicturebox{\let\picturedefinition=\rel@x
   \errhelp=\disabledpictures
   \errmessage{This version of TeX cannot handle pictures.  Sorry.}}
\newhelp\disabledpictures
     {You will get a blank box in place of your picture.}
%
%
%
%
%
%
%
%
%
%
\def\FRONTPAGE{\ifvoid255\else\vfill\penalty-20000\fi
   \gl@bal\pagenumber=1     \gl@bal\chapternumber=0
   \gl@bal\equanumber=0     \gl@bal\sectionnumber=0
   \gl@bal\referencecount=0 \gl@bal\figurecount=0
   \gl@bal\tablecount=0     \gl@bal\frontpagetrue
   \gl@bal\lastf@@t=0       \gl@bal\footsymbolcount=0
   \gl@bal\cn@@false }

\def\papers{\papersize\headline=\paperheadline\footline=\paperfootline}
\def\papersize{\hsize=35pc \vsize=50pc \hoffset=0pc \voffset=1pc
   \advance\hoffset by\HOFFSET \advance\voffset by\VOFFSET
   \pagebottomfiller=0pc
   \skip\footins=\bigskipamount \normalspace }
\papers  
%
%
\newskip\lettertopskip       \lettertopskip=20pt plus 50pt
\newskip\letterbottomskip    \letterbottomskip=\z@ plus 100pt
\newskip\signatureskip       \signatureskip=40pt plus 3pt
\def\lettersize{\hsize=6.5in \vsize=8.5in \hoffset=0in \voffset=0.5in
   \advance\hoffset by\HOFFSET \advance\voffset by\VOFFSET
   \pagebottomfiller=\letterbottomskip
   \skip\footins=\smallskipamount \multiply\skip\footins by 3
   \singlespace }
\def\MEMO{\lettersize \headline=\letterheadline \footline={\hfil }%
   \let\rule=\memorule \FRONTPAGE \memohead }

\def\memodate{\afterassignment\MEMO \date }
\def\memit@m#1{\smallskip \hangafter=0 \hangindent=1in
    \Textindent{\caps #1}}
\def\subject{\memit@m{Subject:}}
\def\topic{\memit@m{Topic:}}
\def\from{\memit@m{From:}}
\def\to{\rel@x \ifmmode \rightarrow \else \memit@m{To:}\fi }
\def\memorule{\medskip\hrule height 1pt\bigskip}  
\def\memohead{\centerline{\fourteenrm MEMORANDUM}}
\newwrite\labelswrite
\newtoks\rw@toks
\def\letters{\lettersize
   \headline=\letterheadline \footline=\letterfootline
   \immediate\openout\labelswrite=\jobname.lab}

\let\letterhead=\rel@x
\def\addressee#1{\medskip\line{\hskip 0.75\hsize plus\z@ minus 0.25\hsize
                               \the\date \hfil }%
   \vskip \lettertopskip
   \ialign to\hsize{\strut ##\hfil\tabskip 0pt plus \hsize \crcr #1\crcr}
   \writelabel{#1}\medskip \noindent\hskip -\spaceskip \ignorespaces }
\def\rwl@begin#1\cr{\rw@toks={#1\crcr}\rel@x
   \immediate\write\labelswrite{\the\rw@toks}\futurelet\n@xt\rwl@next}
\def\rwl@next{\ifx\n@xt\rwl@end \let\n@xt=\rel@x
      \else \let\n@xt=\rwl@begin \fi \n@xt}
\let\rwl@end=\rel@x
\def\writelabel#1{\immediate\write\labelswrite{\noexpand\labelbegin}
     \rwl@begin #1\cr\rwl@end
     \immediate\write\labelswrite{\noexpand\labelend}}
\newtoks\FromAddress         \FromAddress={}
\newtoks\sendername          \sendername={}
\newbox\FromLabelBox
\newdimen\labelwidth          \labelwidth=6in
\def\makelabels{\afterassignment\Makelabels \sendersname=}
\def\Makelabels{\FRONTPAGE \letterinfo={\hfil } \MakeFromBox
     \immediate\closeout\labelswrite  \input \jobname.lab\vfil\eject}
\let\labelend=\rel@x
\def\labelbegin#1\labelend{\setbox0=\vbox{\ialign{##\hfil\cr #1\crcr}}
     \MakeALabel }
\def\MakeFromBox{\gl@bal\setbox\FromLabelBox=\vbox{\Tenpoint
     \ialign{##\hfil\cr \the\sendername \the\FromAddress \crcr }}}
\def\MakeALabel{\vskip 1pt \hbox{\vrule \vbox{
        \hsize=\labelwidth \hrule\bigskip
        \leftline{\hskip 1\parindent \copy\FromLabelBox}\bigskip
        \centerline{\hfil \box0 } \bigskip \hrule
        }\vrule } \vskip 1pt plus 1fil }
\def\signed#1{\par \nobreak \bigskip \dt@pfalse \begingroup
  \everycr={\noalign{\nobreak
            \ifdt@p\vskip\signatureskip\gl@bal\dt@pfalse\fi }}%
  \tabskip=0.5\hsize plus \z@ minus 0.5\hsize
  \halign to\hsize {\strut ##\hfil\tabskip=\z@ plus 1fil minus \z@\crcr
          \noalign{\gl@bal\dt@ptrue}#1\crcr }%
  \endgroup \bigskip }
\newbox\letterb@x
\def\lettertext{\par \vskip\parskip \unvcopy\letterb@x \par }
\def\multiletter{\setbox\letterb@x=\vbox\bgroup
      \everypar{\vrule height 1\baselineskip depth 0pt width 0pt }
      \singlespace \topskip=\baselineskip }
\def\letterend{\par\egroup}
%
%
%
\newskip\frontpageskip
\newtoks\Pubnum   
\newtoks\Pubtype  \let\pubtype=\Pubtype
\newif\ifp@bblock  \p@bblocktrue
\def\PH@SR@V{\doubl@true
\baselineskip=24.1pt plus 0.2pt minus 0.1pt
   \parskip= 3pt plus 2pt minus 1pt }
\def\PHYSREV{\papers\PhysRevtrue\PH@SR@V}

\def\titlepage{\FRONTPAGE\papers\ifPhysRev\PH@SR@V\fi
   \ifp@bblock\p@bblock \else\hrule height\z@ \rel@x \fi }
\def\nopubblock{\p@bblockfalse}
\def\endpage{\vfil\break}
\frontpageskip=12pt plus .5fil minus 2pt

\Pubtype={}
\Pubnum={}
\def\p@bblock{\begingroup \tabskip=\hsize minus \hsize
   \baselineskip=1.5\ht\strutbox\topspace
  \baselineskip\halign to\hsize{\strut ##\hfil\tabskip=0pt\crcr
  \the\Pubnum\crcr\the\date\crcr\the\pubtype\crcr}\endgroup}

\def\title#1{\vskip\frontpageskip \titlestyle{#1} \vskip\headskip }
\def\author#1{\vskip\frontpageskip\titlestyle{\twelvecp #1}\nobreak}

\def\address#1{\par\kern 5pt\titlestyle{\twelvepoint\it #1}}
\def\andaddress{\par\kern 5pt \centerline{\sl and} \address}
\def\abstract{\par\dimen@=\prevdepth \hrule height\z@ \prevdepth=\dimen@
   \vskip\frontpageskip\centerline{\fourteenrm ABSTRACT}\vskip\headskip }

%
%
%

\def\etal{\hbox{\it et al.}}   
\def\\{\rel@x \ifmmode \backslash \else {\tt\char`\\}\fi }
\def\sequentialequations{\rel@x \if\equanumber<0 \else
  \gl@bal\equanumber=-\equanumber \gl@bal\advance\equanumber by -1 \fi }
%

%
\def\journal#1&#2&#3(#4){\begingroup \let\journal=\dummyj@urnal
    \unskip~\sl #1\unskip~\bf\ignorespaces #2\rm
    \unskip,~\ignorespaces #3
    (\afterassignment\j@ur \count255=#4)\endgroup\ignorespaces }
\def\j@ur{\ifnum\count255<100 \advance\count255 by 1900 \fi
          \number\count255 }
\def\dummyj@urnal{%
    \toks@={Reference foul up: nested \journal macros}%
    \errhelp={Your forgot & or ( ) after the last \journal}%
    \errmessage{\the\toks@ }}

\def\topspace{\hrule height 0pt depth 0pt \vskip}

\def\Buildrel#1\under#2{\mathrel{\mathop{#2}\limits_{#1}}}
\def\becomes#1{\mathchoice{\becomes@\scriptstyle{#1}}
   {\becomes@\scriptstyle{#1}} {\becomes@\scriptscriptstyle{#1}}
   {\becomes@\scriptscriptstyle{#1}}}
\def\becomes@#1#2{\mathrel{\setbox0=\hbox{$\m@th #1{\,#2\,}$}%
        \mathop{\hbox to \wd0 {\rightarrowfill}}\limits_{#2}}}

\def\VEV#1{\left\langle #1\right\rangle}

\let\int=\intop         
\def\lsim{\mathrel{\mathpalette\@versim<}}
\def\gsim{\mathrel{\mathpalette\@versim>}}
\def\@versim#1#2{\vcenter{\offinterlineskip
        \ialign{$\m@th#1\hfil##\hfil$\crcr#2\crcr\sim\crcr } }}
\def\big#1{{\hbox{$\left#1\vbox to 0.85\b@gheight{}\right.\n@space$}}}
\def\Big#1{{\hbox{$\left#1\vbox to 1.15\b@gheight{}\right.\n@space$}}}
\def\bigg#1{{\hbox{$\left#1\vbox to 1.45\b@gheight{}\right.\n@space$}}}
\def\Bigg#1{{\hbox{$\left#1\vbox to 1.75\b@gheight{}\right.\n@space$}}}
\def\){\mskip 2mu\nobreak }
%
%
%
\let\sec@nt=\sec
\def\sec{\rel@x\ifmmode\let\n@xt=\sec@nt\else\let\n@xt\section\fi\n@xt}
\def\obsolete#1{\message{Macro \string #1 is obsolete.}}
\def\firstsec#1{\obsolete\firstsec \section{#1}}
\def\firstsubsec#1{\obsolete\firstsubsec \subsection{#1}}
\def\thispage#1{\obsolete\thispage \gl@bal\pagenumber=#1\frontpagefalse}
\def\thischapter#1{\obsolete\thischapter \gl@bal\chapternumber=#1}
\def\splitout{\obsolete\splitout\rel@x}
\def\prop{\obsolete\prop \propto }
\def\nextequation#1{\obsolete\nextequation \gl@bal\equanumber=#1
   \ifnum\the\equanumber>0 \gl@bal\advance\equanumber by 1 \fi}
\def\BOXITEM{\afterassigment\B@XITEM\setbox0=}
\def\B@XITEM{\par\hangindent\wd0 \noindent\box0 }
%
%
%
\def\phyzzx{PHY\setbox0=\hbox{Z}\copy0 \kern-0.5\wd0 \box0 X}
        
\everyjob{\xdef\today{\monthname~\number\day, \number\year}
        \input myphyx.tex }
\message{ by V.K.}
\catcode`\@=12 
%


%
%
\catcode`\@=11
\font\tensmc=cmcsc10
\def\smc{\tensmc}

\def\hcorrection#1{\advance\hoffset by #1 }
\def\vcorrection#1{\advance\voffset by #1 }
\def\wlog#1{}
\newif\iftitle@
\outer\def\heading{\bigbreak\bgroup\let\\=\cr\tabskip\centering
    \halign to \hsize\bgroup\smc\hfill\ignorespaces##\unskip\hfill\cr}
\def\endheading{\cr\egroup\egroup\nobreak\medskip}

\outer\def\endproclaim{\par\ifdim\lastskip<\medskipamount\removelastskip
  \penalty 55 \fi\medskip\rm}
\outer\def\demo#1{\par\ifdim\lastskip<\smallskipamount\removelastskip
    \smallskip\fi\noindent{\smc\ignorespaces#1\unskip:\enspace}\rm
      \ignorespaces}

\hyphenation{man-u-script man-u-scripts ap-pen-dix ap-pen-di-ces}
\hyphenation{data-base data-bases}
\ifx\amstexloaded@\relax\catcode`\@=13
  \endinput\else\let\amstexloaded@=\relax\fi
\newlinechar=`\^^J
\def\eat@#1{}
\def\Space@.{\futurelet\Space@\relax}
\Space@. %
\newhelp\athelp@
{Only certain combinations beginning with @ make sense to me.^^J
Perhaps you wanted \string\@\space for a printed @?^^J
I've ignored the character or group after @.}
\def\futureletnextat@{\futurelet\next\at@}
{\catcode`\@=\active
\lccode`\Z=`\@ \lowercase
{\gdef@{\expandafter\csname futureletnextatZ\endcsname}
\expandafter\gdef\csname atZ\endcsname
   {\ifcat\noexpand\next a\def\next{\csname atZZ\endcsname}\else
   \ifcat\noexpand\next0\def\next{\csname atZZ\endcsname}\else
    \def\next{\csname atZZZ\endcsname}\fi\fi\next}
\expandafter\gdef\csname atZZ\endcsname#1{\expandafter
   \ifx\csname #1Zat\endcsname\relax\def\next
     {\errhelp\expandafter=\csname athelpZ\endcsname
      \errmessage{Invalid use of \string@}}\else
       \def\next{\csname #1Zat\endcsname}\fi\next}
\expandafter\gdef\csname atZZZ\endcsname#1{\errhelp
    \expandafter=\csname athelpZ\endcsname
      \errmessage{Invalid use of \string@}}}}
\def\atdef@#1{\expandafter\def\csname #1@at\endcsname}
\newhelp\defahelp@{If you typed \string\define\space cs instead of
\string\define\string\cs\space^^J
I've substituted an inaccessible control sequence so that your^^J
definition will be completed without mixing me up too badly.^^J
If you typed \string\define{\string\cs} the inaccessible control sequence^^J
was defined to be \string\cs, and the rest of your^^J
definition appears as input.}
\newhelp\defbhelp@{I've ignored your definition, because it might^^J
conflict with other uses that are important to me.}
\def\define{\futurelet\next\define@}
\def\define@{\ifcat\noexpand\next\relax
  \def\next{\define@@}%
  \else\errhelp=\defahelp@
  \errmessage{\string\define\space must be followed by a control
     sequence}\def\next{\def\garbage@}\fi\next}
\def\undefined@{}
\def\preloaded@{}
\def\define@@#1{\ifx#1\relax\errhelp=\defbhelp@
   \errmessage{\string#1\space is already defined}\def\next{\def\garbage@}%
   \else\expandafter\ifx\csname\expandafter\eat@\string
         #1@\endcsname\undefined@\errhelp=\defbhelp@
   \errmessage{\string#1\space can't be defined}\def\next{\def\garbage@}%
   \else\expandafter\ifx\csname\expandafter\eat@\string#1\endcsname\relax
     \def\next{\def#1}\else\errhelp=\defbhelp@
     \errmessage{\string#1\space is already defined}\def\next{\def\garbage@}%
      \fi\fi\fi\next}
\def\famzero{\fam\z@}
\def\arccos{\mathop{\famzero arccos}\nolimits}

\def\arctan{\mathop{\famzero arctan}\nolimits}

\def\cos{\mathop{\famzero cos}\nolimits}

\def\exp{\mathop{\famzero exp}\nolimits}

\def\lim{\mathop{\famzero lim}}

\def\ln{\mathop{\famzero ln}\nolimits}

\def\sec{\mathop{\famzero sec}\nolimits}
\def\sin{\mathop{\famzero sin}\nolimits}

\def\tan{\mathop{\famzero tan}\nolimits}

\def\textfont@#1#2{\def#1{\relax\ifmmode
    \errmessage{Use \string#1\space only in text}\else#2\fi}}
\let\ic@=\/
\def\/{\unskip\ic@}
\def\textfonti{\the\textfont1 }
\def\t#1#2{{\edef\next{\the\font}\textfonti\accent"7F \next#1#2}}
\let\B=\=
\let\D=\.
\def~{\unskip\nobreak\ \ignorespaces}
{\catcode`\@=\active
\gdef\@{\char'100 }}
\atdef@-{\leavevmode\futurelet\next\athyph@}
\def\athyph@{\ifx\next-\let\next=\athyph@@
  \else\let\next=\athyph@@@\fi\next}
\def\athyph@@@{\hbox{-}}
\def\athyph@@#1{\futurelet\next\athyph@@@@}
\def\athyph@@@@{\if\next-\def\next##1{\hbox{---}}\else
    \def\next{\hbox{--}}\fi\next}
\def\.{.\spacefactor=\@m}
\atdef@.{\null.}
\atdef@,{\null,}
\atdef@;{\null;}
\atdef@:{\null:}
\atdef@?{\null?}
\atdef@!{\null!}
\def\srdr@{\thinspace}
\def\drsr@{\kern.02778em}
\def\sldl@{\kern.02778em}
\def\dlsl@{\thinspace}
\atdef@"{\unskip\futurelet\next\atqq@}
\def\atqq@{\ifx\next\Space@\def\next. {\atqq@@}\else
         \def\next.{\atqq@@}\fi\next.}
\def\atqq@@{\futurelet\next\atqq@@@}
\def\atqq@@@{\ifx\next`\def\next`{\atqql@}\else\def\next'{\atqqr@}\fi\next}
\def\atqql@{\futurelet\next\atqql@@}
\def\atqql@@{\ifx\next`\def\next`{\sldl@``}\else\def\next{\dlsl@`}\fi\next}
\def\atqqr@{\futurelet\next\atqqr@@}
\def\atqqr@@{\ifx\next'\def\next'{\srdr@''}\else\def\next{\drsr@'}\fi\next}

\def\textfontii{\the\textfont2 }
\def\{{\relax\ifmmode\lbrace\else
    {\textfontii f}\spacefactor=\@m\fi}
\def\}{\relax\ifmmode\rbrace\else
    \let\@sf=\empty\ifhmode\edef\@sf{\spacefactor=\the\spacefactor}\fi
      {\textfontii g}\@sf\relax\fi}
\def\nonhmodeerr@#1{\errmessage
     {\string#1\space allowed only within text}}
\def\linebreak{\relax\ifhmode\unskip\break\else
    \nonhmodeerr@\linebreak\fi}
\def\allowlinebreak{\relax
   \ifhmode\allowbreak\else\nonhmodeerr@\allowlinebreak\fi}
\newskip\saveskip@
\def\nolinebreak{\relax\ifhmode\saveskip@=\lastskip\unskip
  \nobreak\ifdim\saveskip@>\z@\hskip\saveskip@\fi
   \else\nonhmodeerr@\nolinebreak\fi}
\def\newline{\relax\ifhmode\null\hfil\break
    \else\nonhmodeerr@\newline\fi}
\def\nonmathaerr@#1{\errmessage
     {\string#1\space is not allowed in display math mode}}
\def\nonmathberr@#1{\errmessage{\string#1\space is allowed only in math mode}}
\def\mathbreak{\relax\ifmmode\ifinner\break\else
   \nonmathaerr@\mathbreak\fi\else\nonmathberr@\mathbreak\fi}
\def\nomathbreak{\relax\ifmmode\ifinner\nobreak\else
    \nonmathaerr@\nomathbreak\fi\else\nonmathberr@\nomathbreak\fi}
\def\allowmathbreak{\relax\ifmmode\ifinner\allowbreak\else
     \nonmathaerr@\allowmathbreak\fi\else\nonmathberr@\allowmathbreak\fi}
\def\pagebreak{\relax\ifmmode
   \ifinner\errmessage{\string\pagebreak\space
     not allowed in non-display math mode}\else\postdisplaypenalty-\@M\fi
   \else\ifvmode\penalty-\@M\else\edef\spacefactor@
       {\spacefactor=\the\spacefactor}\vadjust{\penalty-\@M}\spacefactor@
        \relax\fi\fi}
\def\nopagebreak{\relax\ifmmode
     \ifinner\errmessage{\string\nopagebreak\space
    not allowed in non-display math mode}\else\postdisplaypenalty\@M\fi
    \else\ifvmode\nobreak\else\edef\spacefactor@
        {\spacefactor=\the\spacefactor}\vadjust{\penalty\@M}\spacefactor@
         \relax\fi\fi}
\def\newpage{\relax\ifvmode\vfill\penalty-\@M\else\nonvmodeerr@\newpage\fi}
\def\nonvmodeerr@#1{\errmessage
    {\string#1\space is allowed only between paragraphs}}
\def\smallpagebreak{\relax\ifvmode\smallbreak
      \else\nonvmodeerr@\smallpagebreak\fi}
\def\medpagebreak{\relax\ifvmode\medbreak
       \else\nonvmodeerr@\medpagebreak\fi}
\def\bigpagebreak{\relax\ifvmode\bigbreak
      \else\nonvmodeerr@\bigpagebreak\fi}
\newdimen\captionwidth@
\captionwidth@=\hsize
\advance\captionwidth@ by -1.5in
\def\caption#1{}
\def\topspace#1{\gdef\thespace@{#1}\ifvmode\def\next
    {\futurelet\next\topspace@}\else\def\next{\nonvmodeerr@\topspace}\fi\next}
\def\topspace@{\ifx\next\Space@\def\next. {\futurelet\next\topspace@@}\else
     \def\next.{\futurelet\next\topspace@@}\fi\next.}
\def\topspace@@{\ifx\next\caption\let\next\topspace@@@\else
    \let\next\topspace@@@@\fi\next}
 \def\topspace@@@@{\topinsert\vbox to
       \thespace@{}\endinsert}
\def\topspace@@@\caption#1{\topinsert\vbox to
    \thespace@{}\nobreak
      \smallskip
    \setbox\z@=\hbox{\noindent\ignorespaces#1\unskip}%
   \ifdim\wd\z@>\captionwidth@
   \centerline{\vbox{\hsize=\captionwidth@\noindent\ignorespaces#1\unskip}}%
   \else\centerline{\box\z@}\fi\endinsert}
\def\midspace#1{\gdef\thespace@{#1}\ifvmode\def\next
    {\futurelet\next\midspace@}\else\def\next{\nonvmodeerr@\midspace}\fi\next}
\def\midspace@{\ifx\next\Space@\def\next. {\futurelet\next\midspace@@}\else
     \def\next.{\futurelet\next\midspace@@}\fi\next.}
\def\midspace@@{\ifx\next\caption\let\next\midspace@@@\else
    \let\next\midspace@@@@\fi\next}
 \def\midspace@@@@{\midinsert\vbox to
       \thespace@{}\endinsert}
\def\midspace@@@\caption#1{\midinsert\vbox to
    \thespace@{}\nobreak
      \smallskip
      \setbox\z@=\hbox{\noindent\ignorespaces#1\unskip}%
      \ifdim\wd\z@>\captionwidth@
    \centerline{\vbox{\hsize=\captionwidth@\noindent\ignorespaces#1\unskip}}%
    \else\centerline{\box\z@}\fi\endinsert}
\mathchardef\prime@="0230
\def\prime{{{}\prime@{}}}
\def\prim@s{\prime@\futurelet\next\pr@m@s}

\def\,{\relax\ifmmode\mskip\thinmuskip\else\thinspace\fi}
\def\!{\relax\ifmmode\mskip-\thinmuskip\else\negthinspace\fi}
\def\frac#1#2{{#1\over#2}}

\def\:{\nobreak\hskip.1111em{:}\hskip.3333em plus .0555em\relax}
\def\intic@{\mathchoice{\hskip5\p@}{\hskip4\p@}{\hskip4\p@}{\hskip4\p@}}
\def\negintic@{\mathchoice{\hskip-5\p@}{\hskip-4\p@}{\hskip-4\p@}{\hskip-4\p@}}
\def\intkern@{\mathchoice{\!\!\!}{\!\!}{\!\!}{\!\!}}
\def\intdots@{\mathchoice{\cdots}{{\cdotp}\mkern1.5mu
    {\cdotp}\mkern1.5mu{\cdotp}}{{\cdotp}\mkern1mu{\cdotp}\mkern1mu
      {\cdotp}}{{\cdotp}\mkern1mu{\cdotp}\mkern1mu{\cdotp}}}
\newcount\intno@
\def\iint{\intno@=\tw@\futurelet\next\ints@}
\def\iiint{\intno@=\thr@@\futurelet\next\ints@}
\def\iiiint{\intno@=4 \futurelet\next\ints@}
\def\idotsint{\intno@=\z@\futurelet\next\ints@}
\def\ints@{\findlimits@\ints@@}
\newif\iflimtoken@
\newif\iflimits@
\def\findlimits@{\limtoken@false\limits@false\ifx\next\limits
 \limtoken@true\limits@true\else\ifx\next\nolimits\limtoken@true\limits@false
    \fi\fi}
\def\multintlimits@{\intop\ifnum\intno@=\z@\intdots@
  \else\intkern@\fi
    \ifnum\intno@>\tw@\intop\intkern@\fi
     \ifnum\intno@>\thr@@\intop\intkern@\fi\intop}
\def\multint@{\int\ifnum\intno@=\z@\intdots@\else\intkern@\fi
   \ifnum\intno@>\tw@\int\intkern@\fi
    \ifnum\intno@>\thr@@\int\intkern@\fi\int}
\def\ints@@{\iflimtoken@\def\ints@@@{\iflimits@
   \negintic@\mathop{\intic@\multintlimits@}\limits\else
    \multint@\nolimits\fi\eat@}\else
     \def\ints@@@{\multint@\nolimits}\fi\ints@@@}
\def\Sb{_\bgroup\vspace@
        \baselineskip=\fontdimen10 \scriptfont\tw@
        \advance\baselineskip by \fontdimen12 \scriptfont\tw@
        \lineskip=\thr@@\fontdimen8 \scriptfont\thr@@
        \lineskiplimit=\thr@@\fontdimen8 \scriptfont\thr@@
        \Let@\vbox\bgroup\halign\bgroup \hfil$\scriptstyle
            {##}$\hfil\cr}
\def\endSb{\crcr\egroup\egroup\egroup}
\def\Sp{^\bgroup\vspace@
        \baselineskip=\fontdimen10 \scriptfont\tw@
        \advance\baselineskip by \fontdimen12 \scriptfont\tw@
        \lineskip=\thr@@\fontdimen8 \scriptfont\thr@@
        \lineskiplimit=\thr@@\fontdimen8 \scriptfont\thr@@
        \Let@\vbox\bgroup\halign\bgroup \hfil$\scriptstyle
            {##}$\hfil\cr}
\def\endSp{\crcr\egroup\egroup\egroup}
\def\Let@{\relax\iffalse{\fi\let\\=\cr\iffalse}\fi}
\def\vspace@{\def\vspace##1{\noalign{\vskip##1 }}}
\def\aligned{\,\vcenter\bgroup\vspace@\Let@\openup\jot\m@th\ialign
  \bgroup \strut\hfil$\displaystyle{##}$&$\displaystyle{{}##}$\hfil\crcr}
\def\endaligned{\crcr\egroup\egroup}
\def\matrix{\,\vcenter\bgroup\Let@\vspace@
    \normalbaselines
  \m@th\ialign\bgroup\hfil$##$\hfil&&\quad\hfil$##$\hfil\crcr
    \mathstrut\crcr\noalign{\kern-\baselineskip}}
\def\endmatrix{\crcr\mathstrut\crcr\noalign{\kern-\baselineskip}\egroup
                \egroup\,}
\newtoks\hashtoks@
\hashtoks@={#}
\def\format{\crcr\egroup\iffalse{\fi\ifnum`}=0 \fi\format@}
\def\format@#1\\{\def\preamble@{#1}%
  \def\c{\hfil$\the\hashtoks@$\hfil}%
  \def\r{\hfil$\the\hashtoks@$}%
  \def\l{$\the\hashtoks@$\hfil}%
  \setbox\z@=\hbox{\xdef\Preamble@{\preamble@}}\ifnum`{=0 \fi\iffalse}\fi
   \ialign\bgroup\span\Preamble@\crcr}

\def\cases{\left\{\,\vcenter\bgroup\vspace@
     \normalbaselines\openup\jot\m@th
       \Let@\ialign\bgroup$##$\hfil&\quad$##$\hfil\crcr
      \mathstrut\crcr\noalign{\kern-\baselineskip}}

\newif\iftagsleft@
\tagsleft@true
\def\TagsOnRight{\global\tagsleft@false}
\def\tag#1$${\iftagsleft@\leqno\else\eqno\fi
 \hbox{\def\pagebreak{\global\postdisplaypenalty-\@M}%
 \def\nopagebreak{\global\postdisplaypenalty\@M}\rm(#1\unskip)}%
  $$\postdisplaypenalty\z@\ignorespaces}
\interdisplaylinepenalty=\@M
\def\allowdisplaybreak@{\def\allowdisplaybreak{\noalign{\allowbreak}}}
\def\displaybreak@{\def\displaybreak{\noalign{\break}}}
\def\align#1\endalign{\def\tag{&}\vspace@\allowdisplaybreak@\displaybreak@
  \iftagsleft@\lalign@#1\endalign\else
   \ralign@#1\endalign\fi}
\def\ralign@#1\endalign{\displ@y\Let@\tabskip\centering\halign to\displaywidth
     {\hfil$\displaystyle{##}$\tabskip=\z@&$\displaystyle{{}##}$\hfil
       \tabskip=\centering&\llap{\hbox{(\rm##\unskip)}}\tabskip\z@\crcr
             #1\crcr}}
\def\lalign@#1\endalign{\displ@y\Let@\tabskip\centering\halign to \displaywidth
   {\hfil$\displaystyle{##}$\tabskip=\z@&$\displaystyle{{}##}$\hfil
   \tabskip=\centering&\kern-\displaywidth
        \rlap{\hbox{(\rm##\unskip)}}\tabskip=\displaywidth\crcr
               #1\crcr}}
\def\overrightarrow{\mathpalette\overrightarrow@}
\def\overrightarrow@#1#2{\vbox{\ialign{$##$\cr
    #1{-}\mkern-6mu\cleaders\hbox{$#1\mkern-2mu{-}\mkern-2mu$}\hfill
     \mkern-6mu{\to}\cr
     \noalign{\kern -1\p@\nointerlineskip}
     \hfil#1#2\hfil\cr}}}
\def\overleftarrow{\mathpalette\overleftarrow@}
\def\overleftarrow@#1#2{\vbox{\ialign{$##$\cr
     #1{\leftarrow}\mkern-6mu\cleaders\hbox{$#1\mkern-2mu{-}\mkern-2mu$}\hfill
      \mkern-6mu{-}\cr
     \noalign{\kern -1\p@\nointerlineskip}
     \hfil#1#2\hfil\cr}}}
\def\overleftrightarrow{\mathpalette\overleftrightarrow@}
\def\overleftrightarrow@#1#2{\vbox{\ialign{$##$\cr
     #1{\leftarrow}\mkern-6mu\cleaders\hbox{$#1\mkern-2mu{-}\mkern-2mu$}\hfill
       \mkern-6mu{\to}\cr
    \noalign{\kern -1\p@\nointerlineskip}
      \hfil#1#2\hfil\cr}}}
\def\underrightarrow{\mathpalette\underrightarrow@}
\def\underrightarrow@#1#2{\vtop{\ialign{$##$\cr
    \hfil#1#2\hfil\cr
     \noalign{\kern -1\p@\nointerlineskip}
    #1{-}\mkern-6mu\cleaders\hbox{$#1\mkern-2mu{-}\mkern-2mu$}\hfill
     \mkern-6mu{\to}\cr}}}
\def\underleftarrow{\mathpalette\underleftarrow@}
\def\underleftarrow@#1#2{\vtop{\ialign{$##$\cr
     \hfil#1#2\hfil\cr
     \noalign{\kern -1\p@\nointerlineskip}
     #1{\leftarrow}\mkern-6mu\cleaders\hbox{$#1\mkern-2mu{-}\mkern-2mu$}\hfill
      \mkern-6mu{-}\cr}}}
\def\underleftrightarrow{\mathpalette\underleftrightarrow@}
\def\underleftrightarrow@#1#2{\vtop{\ialign{$##$\cr
      \hfil#1#2\hfil\cr
    \noalign{\kern -1\p@\nointerlineskip}
     #1{\leftarrow}\mkern-6mu\cleaders\hbox{$#1\mkern-2mu{-}\mkern-2mu$}\hfill
       \mkern-6mu{\to}\cr}}}
\def\sqrt#1{\radical"270370 {#1}}
\def\dots{\relax\ifmmode\let\next=\ldots\else\let\next=\tdots@\fi\next}
\def\tdots@{\unskip\ \tdots@@}
\def\tdots@@{\futurelet\next\tdots@@@}
\def\tdots@@@{$\mathinner{\ldotp\ldotp\ldotp}\,
   \ifx\next,$\else
   \ifx\next.\,$\else
   \ifx\next;\,$\else
   \ifx\next:\,$\else
   \ifx\next?\,$\else
   \ifx\next!\,$\else
   $ \fi\fi\fi\fi\fi\fi}
\def\text{\relax\ifmmode\let\next=\text@\else\let\next=\text@@\fi\next}
\def\text@@#1{\hbox{#1}}
\def\text@#1{\mathchoice
 {\hbox{\everymath{\displaystyle}\def\textfonti{\the\textfont1 }%
    \def\textfontii{\the\textfont2 }\textdef@@ T#1}}
 {\hbox{\everymath{\textstyle}\def\textfonti{\the\textfont1 }%
    \def\textfontii{\the\textfont2 }\textdef@@ T#1}}
 {\hbox{\everymath{\scriptstyle}\def\textfonti{\the\scriptfont1 }%
   \def\textfontii{\the\scriptfont2 }\textdef@@ S\rm#1}}
 {\hbox{\everymath{\scriptscriptstyle}\def\textfonti{\the\scriptscriptfont1 }%
   \def\textfontii{\the\scriptscriptfont2 }\textdef@@ s\rm#1}}}
\def\textdef@@#1{\textdef@#1\rm \textdef@#1\bf
   \textdef@#1\sl \textdef@#1\it}

\def\textdef@#1#2{\def\next{\csname\expandafter\eat@\string#2fam\endcsname}%
\if S#1\edef#2{\the\scriptfont\next\relax}%
 \else\if s#1\edef#2{\the\scriptscriptfont\next\relax}%
 \else\edef#2{\the\textfont\next\relax}\fi\fi}
\scriptfont\itfam=\tenit \scriptscriptfont\itfam=\tenit
\scriptfont\slfam=\tensl \scriptscriptfont\slfam=\tensl
\mathcode`\0="0030
\mathcode`\1="0031
\mathcode`\2="0032
\mathcode`\3="0033
\mathcode`\4="0034
\mathcode`\5="0035
\mathcode`\6="0036
\mathcode`\7="0037
\mathcode`\8="0038
\mathcode`\9="0039
\def\Cal{\relax\ifmmode\let\next=\Cal@\else
     \def\next{\errmessage{Use \string\Cal\space only in math mode}}\fi\next}
\def\Cal@#1{{\fam2 #1}}
\def\bold{\relax\ifmmode\let\next=\bold@\else
   \def\next{\errmessage{Use \string\bold\space only in math
      mode}}\fi\next}\def\bold@#1{{\fam\bffam #1}}
\mathchardef\Gamma="0000
\mathchardef\Delta="0001
\mathchardef\Theta="0002
\mathchardef\Lambda="0003
\mathchardef\Xi="0004
\mathchardef\Pi="0005
\mathchardef\Sigma="0006
\mathchardef\Upsilon="0007
\mathchardef\Phi="0008
\mathchardef\Psi="0009
\mathchardef\Omega="000A
\mathchardef\varGamma="0100
\mathchardef\varDelta="0101
\mathchardef\varTheta="0102
\mathchardef\varLambda="0103
\mathchardef\varXi="0104
\mathchardef\varPi="0105
\mathchardef\varSigma="0106
\mathchardef\varUpsilon="0107
\mathchardef\varPhi="0108
\mathchardef\varPsi="0109
\mathchardef\varOmega="010A
\font\dummyft@=dummy
\fontdimen1 \dummyft@=\z@
\fontdimen2 \dummyft@=\z@
\fontdimen3 \dummyft@=\z@
\fontdimen4 \dummyft@=\z@
\fontdimen5 \dummyft@=\z@
\fontdimen6 \dummyft@=\z@
\fontdimen7 \dummyft@=\z@
\fontdimen8 \dummyft@=\z@
\fontdimen9 \dummyft@=\z@
\fontdimen10 \dummyft@=\z@
\fontdimen11 \dummyft@=\z@
\fontdimen12 \dummyft@=\z@
\fontdimen13 \dummyft@=\z@
\fontdimen14 \dummyft@=\z@
\fontdimen15 \dummyft@=\z@
\fontdimen16 \dummyft@=\z@
\fontdimen17 \dummyft@=\z@
\fontdimen18 \dummyft@=\z@
\fontdimen19 \dummyft@=\z@
\fontdimen20 \dummyft@=\z@
\fontdimen21 \dummyft@=\z@
\fontdimen22 \dummyft@=\z@
\def\fontlist@{\\{\tenrm}\\{\sevenrm}\\{\fiverm}\\{\teni}\\{\seveni}%
 \\{\fivei}\\{\tensy}\\{\sevensy}\\{\fivesy}\\{\tenex}\\{\tenbf}\\{\sevenbf}%
 \\{\fivebf}\\{\tensl}\\{\tenit}\\{\tensmc}}
\def\dodummy@{{\def\\##1{\global\let##1=\dummyft@}\fontlist@}}
\newif\ifsyntax@
\newcount\countxviii@
\def\newtoks@{\alloc@5\toks\toksdef\@cclvi}
\def\nopages@{\output={\setbox\z@=\box\@cclv \deadcycles=\z@}\newtoks@\output}
\def\syntax{\syntax@true\dodummy@\countxviii@=\count18
\loop \ifnum\countxviii@ > \z@ \textfont\countxviii@=\dummyft@
   \scriptfont\countxviii@=\dummyft@ \scriptscriptfont\countxviii@=\dummyft@
     \advance\countxviii@ by-\@ne\repeat
\dummyft@\tracinglostchars=\z@
  \nopages@\frenchspacing\hbadness=\@M}
\def\wlog#1{\immediate\write-1{#1}}
\catcode`\@=\active

\def\ln{\hbox{ln}}

\font\bfgreek=cmmib10 scaled\magstep1

\hcorrection{1truein}
\vcorrection{1 truein}


\def\bmath{\fam1\bfgreek\textfont1=\bfgreek}

\def\rf#1{$^{#1}$}

\def\ltwid{\raise.3ex\hbox{$<$\kern-.75em\lower1ex\hbox{$\sim$}}}
\def\gl{\raise.5ex\hbox{$>$}\kern-.8em\lower.5ex\hbox{$<$}}
\def\gtwid{\raise.3ex\hbox{$>$\kern-.75em\lower1ex\hbox{$\sim$}}}

\def\3he{$^3\text{He}$}
\def\4he{$^4\text{He}$}

\def\etal{{\it et al.}}


\def\VEC#1{\bold{#1}}
\newcount\refnum
\global\refnum=0
\def\bib{\global\advance\refnum by 1\item{\the\refnum.}}
\def\a0{\text{ \AA}}

\def\pa{\partial}
\def\vecr{\VEC r}

\def\rt  {({\VEC r},t)}
\def\sqr#1#2{{\vcenter{\vbox{\hrule height.#2pt
     \hbox{\vrule width.#2pt height#1pt \kern#1pt
     \vrule width.#2pt}
     \hrule height.#2pt}}}}
 
\def\bm#1{\hbox{$\bmath #1$}}
\HOFFSET=0.375truein

\tolerance 2000
\nopublock

\titlepage
\title{\bf Dynamics of Phase Separation of Crystal Surfaces}
\author{Fong Liu\rf{1} and Horia Metiu\rf{2}}
\address{\rf1 Institute for Theoretical Physics and Center for
Quantized\break Electronic Structures, University of
California, Santa Barbara, CA 93106}
\address{\rf2Department of Chemistry and Physics,
University of California,\break Santa Barbara, CA 93106}
\vfil
\abstract
\noindent We investigate the dynamical evolution of a
thermodynamically unstable crystal surface into
a hill-and-valley structure.
We demonstrate that, for quasi one-dimensional ordering, the
equation of motion maps exactly to the modified Cahn-Hilliard
equation describing spinodal decomposition. Orderings in two
dimensions follow the dynamics of continuum clock models.
Our analysis emphasizes the importance of crystalline
anisotropy and the interaction between phase boundaries in
controlling the long time dynamics.  We establish that the
hill-and-valley pattern coarsens logarithmically in time for
quasi one-dimensional orderings.
For two-dimensional orderings, a power-law growth $L(t)\sim t^n$
of the typical pattern size
is attained with exponent $n\approx 0.23$ and $n\approx 0.13$,
for the two ordering mechanisms dominated by
evaporation-condensation and by surface-diffusion respectively.


\bigskip\noindent
PACS numbers: 68.35.-p, 64.60.Cn
\vfil
\endpage

\chapter{INTRODUCTION}
It was first realized by Curie and Gibbs\rlap,\Ref\gibbs{
P. Curie, \journal Bull. Soc. Min. France &8&145(1885); J.W. Gibbs,
\journal Trans. Conn. Acad. &3&343(1878).}
over a century ago, that the equilibrium shape of a small crystal is
 determined by minimizing its total surface free energy.
 The explicit
solution of this minimization problem was given by Wulff in
1901\rlap,\Ref\wulff{G. Wulff, \journal Z. Krist. &34&449(1901). For a
recent review, see M. Wortis in {\sl Chemistry and Physics of Solid
Surfaces}, vol.7, edited by R. Vanselow and R. Howe (Springer, Berlin,
1988).}  in terms of the famous Wulff construction using a
polar plot of the surface free energy density $\gamma$.
The surface free energy density of
 a crystalline solid is, in general,
anisotropic, due to the underlying crystalline symmetry. As a result,
an equilibrium crystal shape (ECS) may exhibit a complex
topography. For sufficiently strong anisotropy,
facets, edges, or even corners can been seen on an ECS. Meanwhile,
certain orientations  become energetically unfavorable and
are entirely excluded from the ECS. These considerations have
important implications for the stability of a large crystal
surface of constant orientation. In fact, a surface with an
orientation not appearing on the ECS can break up into
hill-and-valley structures, so as to lower its total
surface free energy in spite of a larger surface area. This process,
termed variously as thermal faceting or thermal etching in the
metallurgy literature, has been studied for many
decades\rlap.\REFS\moore{A.J.W. Moore in {\sl
Metal Surfaces: Structures, Energetics and Kinetics},
(American Soc. Metals, Metals Park, 1963).}\REFSCON\flytzani{
M. Flytzani-Stephanopoulos and L.D. Schmidt, \journal Progress
in Surface Science, &9&83(79).}\refsend

The static problem of the thermodynamic stability of a
 crystal surface was first thoroughly investigated by
Herring\rlap,\Ref\herring{C. Herring, \journal Phys. Rev. &82&87(51); C.
Herring in {\sl Structure and Properties of Solid Surfaces},
edited by R. Gomer and C.S.  Smith (University of Chicago Press,
Chicago, 1953).} utilizing a geometric analysis of the $\gamma$-plot.
His criterion of stability, simply put, states that a flat surface
is thermodynamically stable with respect to formation of
hill-and-valley patterns if, and only if, its orientation is present
in the ECS.  It was later realized by Cabrera\Ref\cabrera{
N. Cabrera in {\sl Symposium on Properties of Surfaces},
(American Society for Testing and Materials, Philadelphia, 1963).}
 that Herring's analysis can be equivalently formulated in a
 language commonly used in describing the
thermodynamics of unstable coexisting volume phases.
Cabrera's approach suggests a possible close analogy of the
problem of surface stability to that of bulk phase separation of
binary systems.
This line of thought was recently resurrected by Stewart and
Goldenfeld\rlap,\Ref\stewart{J. Stewart and N. Goldenfeld,
\journal Phys. Rev. A  &46&6505(92).}
who extended the above analogy to dynamic problems and derived
an equation of motion for the crystal surface.

The dynamical evolution of crystal surface morphology was
first studied by Mullins in a series of pioneering
papers\rlap.\REFS\grooving{
W.W. Mullins, \journal J. App. Phys. &28&333(1957); \journal
Acta Metall. &6&414(58); \journal J. App. Phys.
&30&77(59).}\REFSCON\etch{W.W. Mullins, \journal Phil. Mag.
&6&1313(61).}\refsend
In particular, Mullins\refmark{\etch}
considered the growth kinetics of a single, linear
facet on a crystal-vapor interface under two distinct
transport mechanisms: evaporation-condensation,
and surface-diffusion.
 Approximate linear evolution equations were derived,
 under the assumption of a nearly flat
surface and of isotropic surface energy and transport coefficients.
Mullins then proceeded to show, based on approximate analytic
solution of the linear equations, that the typical facet size $L(t)$
should increase with time as a power law, i.e.,  $L(t)\sim t^n$.
The exponent was predicted to be $n=\frac{1}{2}$
($\frac{1}{4}$) for the evaporation-condensation
 (surface-diffusion)  mechanism.
In their recent paper, Stewart and Goldenfeld\refmark{\stewart}
performed a linear stability analysis of the development of
instabilities in the early stages. The propagation of facets,
as well as the role of elastic interactions on the
phase diagram were also investigated.
On the experimental side, many efforts have
been made to corroborate Mullins' growth laws,
but the outcome was either inclusive, or the measured facet
growth were much slower than that predicted
by Mullins\rlap.\Ref\exper{Relevant experiments have been
reviewed in  Ref.\moore\ and Ref.\flytzani. See also, for example,
R.J. Phaneuf, \etal, \journal Phys. Rev. Lett. &67&2986(91).}

In this paper, we investigate the dynamics of an unstable
crystal surface, focusing on the nonlinear evolution at late stages.
The anisotropy in the surface free energy density is taken into
account explicitly, which provides the primary driving force
for phase separation. By introducing a proper order parameter,
the close analogy of the current problem with the
dynamics of first-order phase transitions\Ref\gunton{
J.D. Gunton, M. San Migual and P.S. Sahni in {\sl Phase
Transitions and Critical Phenomena}, edited by C. Domb and
 J.L. Lebowitz (Academic, New York, 1983).} is further elucidated.
Specifically, we demonstrate that, for quasi-one dimensional
ordering, the equation of motion for the order parameter
maps exactly to a modified form of the Cahn-Hilliard
equation\Ref\cahn{J.W.Cahn
and J.E. Hilliard, \journal J. Chem. Phys. &28&258(58).}
describing spinodal decomposition\rlap.\Ref\langerbaron{
J.S. Langer, M. Bar-on and H.D. Miller, \journal Phys. Rev. A.
&11&1417(75).} On the other hand, ordering dynamics
in two dimensions follow that of the continuum clock
models\rlap.\Ref\clockmodel{Fong Liu
and G.F. Mazenko, \journal Phys. Rev. B. &47&2866(93).}
Differing from a usual binary system, the order parameter
used here has intuitive geometrical meaning: it probes the
local inclination of the surface. This establishes an
interesting correspondence between the geometrical singularities
of the surface ( e.g., corners and edges) and the so-called
topological defects (e.g., vortices and interfaces)
commonly encountered in studying  bulk phase separations.
The roles of these defects should therefore be emphasized, since
the interaction and evolution of topological defects are believed
to control the asymptotic ordering dynamics at long times.
Our analysis then shows that the typical size of the
hill-and-valley structures grows logarithmically in time,
$L(t) \sim \ln t$, for quasi one-dimensional systems,
for both the evaporation-condensation and
the surface-diffusion mechanism.
In two dimensions,  the growth law is well-fitted by
a power-law of the form $L(t)\sim t^n$, with  $n \approx 0.23$
for the evaporation-condensation mechanism, and $n \approx 0.13$ for
the surface-diffusion mechanism.
These results provide a plausible explanation for the
slow growth observed in experiments.

\chapter{EQUATIONS OF MOTION}
Consider a macroscopic crystal surface either in vacuum or
in local thermodynamic equilibrium with its own vapor phase.
The position of the interface can be represented
by a height function $z(\vecr,t)$, where $\vecr=(x,y)$,
in a coordinate system which is pinned to the bulk crystal
substrate. This representation is unique, provided that the surface
has no overhangs.
Assuming that the evolution of surface structure does
not involve volume changes of the bulk phases, the
relevant thermodynamic potential is the total surface free energy
$$ \Omega = \int \gamma( \bm n) \, dA \, ,    \eqn\freeen $$
the minimization of which drives the surface evolution.
Here the integration is over the crystal surface and
the surface free energy density $\gamma$ depends on the orientation
of the surface represented by its unit normal
$$  \bm n = ( -p, -q, 1)/\sqrt{1+p^2+q^2}  \eqno\eq $$
where $p=\pa z/\pa x$, $q=\pa z/\pa y$ are two independent
``slopes'' of the surface.
To make connection with phase transition kinetics, it is
beneficial to introduce a 2-vector (or equivalently, a complex)
order parameter $\bm\psi(\vecr,t)=\left(p(\vecr,t), q(\vecr,t)
\right)$, representing the local orientation of the surface.
When considering the evolution of an extended surface,
it is also appropriate to recast \freeen\ into an integration over
 the substrate
$$ \Omega=\int \beta(\bm\psi) \, d^2\vecr  \eqn\intsub $$
using the projected free energy density
$ \beta(\bm\psi) \equiv \gamma(\bm\psi)\sqrt{1+|\bm\psi|^2}$.
For the sake of simplicity, we have tentatively
neglected elastic contributions caused by surface stresses.

In thermodynamic equilibrium, the geometry of a small crystal is
determined uniquely (up to an overall scale factor) by the
functional form of $\gamma(\bm n)$ via the famous Wulff
construction\rlap.\refmark{\wulff}
The convex surface obtained from the Wulff construction minimizes
the total free energy \freeen, and gives the equilibrium crystal
shape. Since $\gamma$ is anisotropic, the resulting ECS
may contain edges and corners. These ideally
sharp objects correspond to discontinuities in the surface
orientation and lead to mathematical singularities
which are difficult to handle in kinetics.
 We can surmount this difficulty by
smoothing out the corners and edges through the introduction
of energy penalties for rapid changes of surface orientation in space.
This treatment is pertinent, since it recaptures the
physically relevant edge and corner energies that were
originally ignored in the Wulff-type
considerations.  The above regularization procedure is reminiscent of
the approach by which  more commonly seen defects
(vortex lines, domain
walls, monopoles, etc.) are treated. In fact, as we shall see,
the edges and corners we encounter here have a direct correspondence
with those defects.
In the rest of the paper,  the terms edge and corner
represent these regularized structures.

The orientation of the surface, upon which the surface free energy
density depends, is characterized by the vector order parameter
$\bm\psi=(p,q)$.
The above discussion suggests that the surface free energy density
also depends on the change of orientation, which is described by the
mean surface curvature $\kappa\approx -\nabla\cdot
  \bm\psi= -(\pa p/\pa x+\pa q/\pa y)$. Expanding $\gamma$  in terms
of the curvature leads to
$$ \gamma(\bm\psi, \nabla\bm\psi) =\gamma_{0}(\bm\psi) +
\gamma_2(\bm\psi) (\nabla\cdot\bm\psi)^2 + \gamma_4(\bm\psi)
(\nabla\cdot\bm\psi)^4+ ... \eqn\gexp $$
where $\gamma_0(\bm\psi)$ is the homogeneous part of the surface free
energy density used in the construction of ECSs.
For symmetry reasons we have retained only terms even in the
gradients.  Eq. \gexp\ should be regarded as the simplest
phenomenological approximation in the spirit of Landau.
The  projected free energy density in \intsub\ can be
similarly written  as
$$ \beta(\bm\psi, \nabla\bm\psi)= \beta_0(\bm\psi) +\frac{a}{2}
     (\nabla\cdot\bm\psi)^2  \eqno\eq $$
where we have further truncated the expansion and retained only
the leading gradient term  with a constant coefficient.
The constant $a$ serves as the regularization cutoff,
determining the equilibrium core size of defects.

Changes of surface topography in time are achieved by
matter transport through the system, driven by  chemical potential
differences.  In our system, the chemical potential relates
to local surface geometry through a modified  Gibbs-Thompson
relation, taking into account the {\it anisotropy} in the surface
free energy density. We start from the definition
 $$\mu-\mu_0 =\omega_0 \delta \Omega /\delta z
             =\omega_0 \frac{\pa}{\pa
   z}\int\beta(\bm\psi,\nabla\bm\psi)d^2\vecr     \eqn\chemdef $$
where $\mu$ is the chemical potential just beneath the
surface, $\mu_0$ is that beneath a planar surface in equilibrium
and $\omega_0$ is the molecular volume of the solid phase.
Expression \chemdef\ is a generalization from the well-known
Gibbs-Thompson formula $\mu-\mu_0=2\omega_0\gamma\kappa$ for
an isotropic surface to anisotropic situations.
Now a straightforward variational calculation\refmark{\stewart}
yields
$$ \mu-\mu_0=\omega_0 a\nabla^2 \left( \nabla\cdot\bm\psi
     \right) -\omega_0 \left(
    \frac{\pa}{\pa x}\frac{\pa \beta_0}{\pa p}
     +\frac{\pa}{\pa y}\frac{\pa \beta_0}{\pa q}
     \right)   \eqn\chemfirst $$
which we can rewrite using the equivalent complex representation
of $\bm\psi=p+iq$.  Realizing that
 $$ \frac{\pa\beta_0(\bm\psi,\bm\psi^*)}{\pa \bm\psi^*}
   =\frac{1}{2}\left(\frac{\pa \beta_0(p,q)}{\pa p},
  \frac{\pa\beta_0
     (p,q)}{\pa q}\right), \eqno\eq $$
we obtain a more concise form for the chemical potential
$$ \mu\left( \{\bm\psi(\vecr,t)\} \right) -\mu_0=
\omega_0 \, \nabla \cdot \left( a\nabla^2 \bm\psi
      -2 \frac{\pa \beta_0(\bm\psi,\bm\psi^*)}{\pa \bm\psi^*}
        \right).   \eqn\chem $$
where it should be warned that we have used interchangeably
both the vector and complex notations of $\psi$.
Without loss of generality, we further absorb the extra factor
of two into $\beta_0$ and set $\omega_0=1$ hereafter.

Various transport mechanisms contribute to the equilibration of
the crystal surface, including evaporation-condensation,
surface diffusion, volume diffusion, as well as plastic flow.
The former two mechanisms were carefully considered by
Mullins\refmark{\grooving} and,
more recently, by Stewart and Goldenfeld\rlap.\refmark{\stewart}
It was shown that their contributions to the normal growth velocity
of the surface has the following form:
$$ v_n = -A \, \mu  + B \,  \nabla^2 \mu   \eqn\velo $$
where the first term on the r.h.s. is due to evaporation-condensation
while the second term comes from surface diffusion. Here
$A$ and $B$ are constants related to material properties
as well as to parameters of the bulk equilibrium state.
In deriving \velo\ it has
been assumed that the transport coefficients are isotropic.
Using
$$  \frac{\pa \bm\psi}{\pa t}=\nabla \frac{\pa z}{\pa t}
    =\nabla \left(v_n \sqrt{1+|\bm\psi|^2}\right),  \eqno\eq $$
along with \chem, equation \velo\ leads immediately to the
deterministic equation of motion for the vector order parameter:
$$ \frac{\pa \bm\psi(\vecr,t)}{\pa t} = \left(-A+B\nabla^2\right)
\nabla\left[\nabla\cdot\left(a\nabla^2\bm\psi(\vecr,t)
 -\frac{\pa\beta_0(\bm\psi,\bm\psi^*)}{\pa \bm\psi^*} \right)\right] .
  \eqn\eqmo $$
In the above derivations we have made the assumption that the
surface has small angles of inclination with respect to the
substrate so that $|\bm\psi| \ll 1$.
A similar equation, though in a different form, was derived
recently by Stewart and Goldenfeld\rlap.\refmark{\stewart}
As will become apparent later, our formulation displays more
conveniently the correspondence with phase transition kinetics.

\chapter{QUASI ONE-DIMENSIONAL ORDERING}
In this section, we investigate the dynamics of quasi one-dimensional
ordering of crystal surfaces. One example is the
linear thermal faceting problem examined by Mullins\refmark{\etch}
in which, by focusing on a cross-section perpendicular
to the facet axis, the surface profile is a function of
one coordinate only $z(\vecr,t)=z(x,t)$. This geometry
is relevant in general to systems with strong preference to order
in one particular direction of the surface\rlap.\Ref\oneexp{
R.J. Phaneuf \etal, in Ref.\exper; R.J. Phaneuf \etal,
{\sl Abs. Mat. Res. Soc.}, page 81, 1992.}
As another example of a quasi 1D geometry, we
can imagine the 1D surface profile as being the perimeter of
islands or steps epitaxially grown on a crystal
surface\rlap.\REFS\comsa{
T. Michely and G. Comsa, \journal Surf. Sci. &256&217(91);
T. Michely, T. Land, U. Littmark and G. Comsa, \journal
Surf. Sci. &272&204(92).}\REFSCON\bartelt{For a review on
surface morphology of epitaxially grown surfaces,
see E.D. Williams, and N.C. Bartelt, \journal Science,
&251&393(91).}\refsend

Since the order parameter is now a scalar, we obtain from \eqmo\ that
$$ \frac{\pa \psi(x,t)}{\pa t}= -A \frac{\pa^2}{\pa x^2}
      \left( a \,  \frac{\pa^2\psi}{\pa x^2}
        -\frac{d\beta_0(\psi)}{d\psi}\right)      \eqn\criticevap $$
for the evaporation-condensation mechanism and
$$  \frac{\pa \psi(x,t)}{\pa t}= B \frac{\pa^4}{\pa x^4}\left( a  \,
       \frac{\pa^2\psi}{\pa x^2} -\frac{d\beta_0(\psi)}
    {d\psi} \right)   \eqn\criticsurf $$
for the surface-diffusion mechanism.
Equation \criticevap\ is identically the Cahn-Hilliard
equation describing spinodal decomposition of a binary system.
We conclude therefore that when the evaporation-condensation
process is the dominant mechanism for matter transport, the
dynamics of the crystal surface follow that of the
one-dimensional spinodal decomposition.
In the  more interesting
case of surface ordering by surface-diffusion, the equation of
motion \criticsurf\ also shows close resemblance to the
Cahn-Hilliard equation in that the outer Laplacian operator is
replaced by a fourth-order derivative. Both cases fall into the
 category of dynamics with conserved order parameters.

Before we proceed further with the dynamics, it is beneficial to
examine first some equilibrium properties of the crystal surface.
Most of the insights can be gained by considering the following
paradigmatic example of an anisotropic  surface free energy density
with four-fold symmetry
$$\gamma_0(\theta) =\sigma_0(1+\epsilon \cos 4\theta) \eqn\odsf $$
where $\sigma_0$ is a constant  and $\theta$ is the angle between the
surface normal and the $(11)$-direction of the crystal. Here the
anisotropy strength $\epsilon \in [0,1]$, which should contain
implicit temperature dependence for a real
system, is the main control parameter
in our considerations.  The equilibrium crystal
shape corresponding to \odsf\ can be determined
by minimizing \freeen\ via the Wulff
construction\rlap,\refmark{\wulff}
the result of which is displayed in Fig.1.
Due to symmetry, we can focus on one quarter of the surface
with $-\frac{\pi}{4} \le \theta \le \frac{\pi}{4}$. It is
found that there is a critical value of anisotropy
$\epsilon_c=\frac{1}{15}$  below which the equilibrium
crystal shape is smooth, containing all possible orientations
$\theta \in [-\frac{\pi}{4},\frac{\pi}{4}]$ (see Fig.1.(a)).
This picture is drastically changed when $\epsilon >
\epsilon_c$. Now sharp corners develop on the ECS and
consequently there exist a finite range of orientations which are
prohibited from appearing in the ECS (Fig.1.b and Fig.1.c).
These absent orientations, between the interval $[\theta^{-},
\theta^{+}]$, are known as the missing angles.
The missing angles can either be read off directly from the
ECS, or can be determined rather easily using the
Frank diagram\rlap.\Ref\frankd{F.C. Frank, in Ref.\moore;
see also J.L. Meijering, \journal Acta. Metall. &11&847(63).}
The Frank diagram is by definition the ploar plot of the
inverse of surface free energy density $\gamma^{-1}(\theta)$. It
can be shown that the missing angles correspond to portions of
the Frank diagram which  are not globally-convex. In our case,
this argument leads trivially to the following condition
determining $\theta^+, \theta^-$:
 $$ \frac{d}{d\theta}\left( \gamma^{-1}(\theta)
    \cos\theta \right)\bigg|_{\theta^+,\theta^-}
    =0,    \eqno\eq $$
which, with \odsf, gives
$$ \theta^{+}=-\theta^{-}=\arccos \frac{1}{\sqrt{6}}\left[
 1+ \left( \frac{5\epsilon+3}{2\epsilon}\right)^{1/2} \right]^{1/2}
  .     \eqno\eq $$
For example, in Fig.1(b) with $\epsilon=0.2$, we obtain
$\theta^+=-\theta^- \approx 33.6^{\circ}$.
In the extreme case of $\epsilon=1$, we have $\theta^{+}=-\theta^{-}
=45^{\circ}$. The equilibrium crystal shape is therefore a
completely faceted square with only four possible
orientations (Fig.1.c).

\FIG\one{Equilibrium crystal shapes calculated from
   the surface free energy density \odsf\ using the Wulff
   construction, with anisotropy strengths (a) $\epsilon=0.02$,
   (b) $\epsilon=0.2$, and (c) $\epsilon=1.0$.
   The corresponding projected free energy profiles
   $\beta_0(\psi)$ as function of the order parameter $\psi$
   are shown on the right. ECSs (b) and (c) contain missing
   orientations.}

\section{Critical Quench}
We now address the question: If we prepare an infinite surface
perpendicular to the crystal $(11)$-direction (taken as the
$z$-axis), how does the interface evolve dynamically?
Herring's argument\refmark{\herring} suggests that if this
surface orientation ($\theta=0$) does not appear in
the equilibrium crystal shape, the surface will be unstable
against the breaking up into hill-and-valley patterns.
In our language, we shall find that this situation corresponds
to the usual phase separation process with a critical quench.

To examine the dynamical stability of the surface, it suffices to
study the projected free energy landscape
$$ \beta_0(\psi)= \sigma_0 (1+\psi^2)^{1/2}
    \left[ 1-\epsilon +2\epsilon\left(\frac{1-
    \psi^2}{1+\psi^2}\right)^2\right]   \eqn\fland $$
obtained from \odsf\ using $\psi=\tan \theta$.
It is worth noting that \fland\ incorporates two effects:
the orientation dependence of the surface free energy,
and the free energy penalty for tilting the surface away from
the original orientation. This last factor leads to interesting
features to be discussed when we consider off-critical quenches.
The profiles of \fland\  for different anisotropy strength
$\epsilon$ are shown on the right in
Fig.1.  $\beta_0(\psi)$ develops a degenerate double-well
 feature when $\epsilon>\epsilon_c$, with minima at
$\psi^{+}=\tan\theta^{+}$ and $\psi^{-}=\tan\theta^{-}$.
Quantity $\epsilon_c$ is the same critical anisotropy at
which the ECS develops sharp corners. It is given by the
stability condition
   $$ \frac{\pa^2 \beta_0}{\pa\psi^2}\bigg|_{\epsilon=\epsilon_c,
\psi=0}=0. \eqno\eq $$
We therefore conclude that when $\epsilon>\epsilon_c$,
the original crystal surface will phase separate into two
distinct homogeneous phases corresponding to stable orientations
$\theta^-$ and $\theta^{+}$.  It is interesting
to point out that these two ``selected'' orientations $\theta^-$ and
$\theta^+$, corresponding to minima of $\beta_0(\psi)$ rather
than $\gamma_0(\psi)$,  are in some sense {\it marginally} stable.
Those directions $\theta^+ <\theta <\frac{\pi}{4}$, although having
smaller surface free energy density, and hence
in some sense being more stable, are not selected due to the
energy penalty for tilting mentioned above.
In Fig.2(a), we plot evolution patterns of an unstable surface
calculated from numerical solutions of \criticsurf\ under
surface diffusion mechanism.  Equation \criticsurf\ can
be made dimensionless by setting the
constants $a, B, \sigma_0$ to unity through proper rescaling of
time and length scales. The resulting dimensionless equation,
containing only $\epsilon$ as a control parameter, is
solved using explicit finite difference method starting from an
initial surface with small amplitude random fluctuations.
As can be seen from Fig.2(a),
typical phase-separated domains take the
form of flat ``facets'' bounded by ``sharp'' edges.

The analogy of our system to spinodal decomposition points to
the following correspondences:
the stable orientations $\psi^+, \psi^-$ correspond to stable
volume phases in bulk phase separations; the edges are
identified with with the interfaces separating different
volume phases.
Strictly speaking, due to our regularization, the
phase-separated surface can not have ideally sharp edges.
The sharpness of the edge, analogous to the thickness of
the interface, is controlled by the regularization parameter
$a$\rlap.\Ref\pego{The Cahn-Hilliard equation can be
regarded as a regularization of the classical Stefan problem
of two coexisting bulk phases. See R.L. Pego, \journal
Proc. Roy. Soc. Lond. A&422&261(89).} In the limit
$a\rightarrow 0$, ideal singularities are recovered.
Similarly, since we have not incorporated cusps in the surface
free energy (see discussions below), the phase-separated
surfaces will not be truly flat facets on atomic scales either.
We shall, however, use the term facet loosely, as denoting regions
of surface whose orientation has saturated to one of the stable
directions. And the process of phase separation will therefore
frequently be referred to as faceting.

\FIG\two{Surface profiles $z(x,t)$ following the dynamical
evolution of an unstable surface into hill-and-valley structures.
(a) Initial surface perpendicular to the $(11)$-direction (critical
quench, $\theta_v=0$). Curves correspond, from bottom up, to
times $t=2$, $3$, $20$ and $80$. Profiles have been displaced
vertically for illustration proposes.
(b) Initial surface tilted $\theta_v=10^{\circ}$ from the
$(11)$-direction (off-critical quench). Times are from bottom
up $t=6$, $12$, $20$, and $80$. In both (a) and (b), $\epsilon=0.2$.}

Results known\refmark{\gunton, \langerbaron}
for the one-dimensional (albeit somewhat artificial)
spinodal decomposition  can be applied
directly to the dynamics of faceting. First of all, an early stage
with exponential growth of the order parameter amplitude is expected.
During this stage, long wavelength fluctuations of the surface
get amplified. However, the typical size of facets will remain
roughly the same, about the order of the most unstable wavelength
of the instability given by
$$ \lambda_c \sim (a/\sigma_0)^{1/2}.  \eqno\eq $$
At later times, the orientations of the phase-separated domains
have almost saturated to their stable values. Subsequent evolutions
of the surface will then be governed by the nonlinear interaction
and motion of the edges. The average size of facets can increase,
only at the expense of smaller facets disappearing through edge-edge
annihilation. Previous investigations\REFS\langaop{
J.S. Langer, \journal Ann. Phys. &65&53(71).}\REFSCON\kawa{
T. Kawakatsu and T. Nunakata, \journal Prog. Theor. Phys.
&74&11(85).}\refsend
of the 1D coarsening dynamics of the Cahn-Hilliard
equation \criticevap\ all reached the conclusion that the
typical domain size grows logarithmically in time during
the late stages,
$$   L(t) \sim \ln \, t  .  \eqn\loggr $$
Similar arguments can be developed for the dynamics
of \criticsurf\ as well\rlap.\Ref\spectrumcom{Fong Liu,
unpublished.} Thereby the conclusion is that the growth
law \loggr\ remains valid for both the evaporation-condensation
and the surface-diffusion mechanisms.
This slow logarithmic growth, coming from the rather
ineffective one-dimensional transport, is consistent with
the experimental observations\refmark{\oneexp}
where very small growth exponents were obtained in contrast to
the predictions of  Mullins' theory.

Mullins' growth law was based upon considerations of single-facet
growth using linear equations of motion. To scrutinize his
analysis, we linearize our equation of motion \criticsurf\
for small order parameter and obtain
 $$ \frac{\pa\psi(x,t)}{\pa t}= -15\sigma_0 B(\epsilon_c
    -\epsilon)\frac{\pa^4\psi}{\pa x^4} +aB\frac{\pa^6\psi}{
    \pa x^6} .   \eqn\lsa  $$
The equation derived by  Mullins is recovered if we
discard the second contribution from the r.h.s.,
and only if $\epsilon<\epsilon_c$ so that the
surface free energy density is only weakly anisotropic.
The latter constraint limits the utility of Mullins'
equation to situations such as the linear relaxation of
{\it stable} or {\it metastable} surfaces only.
Now if we consider the relaxation of a long-wavelength
sinusoidal perturbation applied to an isotropic surface
($\epsilon=0$), by writing $\delta\psi \sim \exp(
 \omega_k t + ikx)$, we obtain from \lsa\ the dispersion relation
$\omega \prop -k^4$, a well-known result derived by Mullins.
Thus it is apparent that the Mullins' analysis in its original
form does not adequately describe the homogeneous breaking
up of unstable surfaces since it neglected two key elements of
the problem: the crystalline anisotropy and the defect (edge)
energy. The first factor, as we have seen, provides the
crucial nonlinear driving force, while the second factor influences
the dynamics of phase boundaries.

\section{Off-critical Quench}
So far we have considered a surface whose normal coincides
with the $(11)$-direction of the crystal.
If on the other hand, the initial surface has its normal
slightly mis-oriented away
from the $(11)$-direction by an angle $\theta_v$ (assumed positive),
we will encounter the situation of an off-critical quench.
Let us still choose $z$-axis normal to the initial surface.
Since $\gamma_0$ is measured with respect to the $(11)$-direction
rather than the $z$-axis,
the projected free energy density now depends on the
tilting angle:
 $$ \beta_0(\psi)= \frac{\gamma_0(\theta+\theta_v)}{\cos\theta}
 \eqn\ftil $$
where $\theta=\arctan\psi$ is measured with respect to the
$z$-axis.  As $\epsilon>\epsilon_c$ and for small enough
$\theta_v$ (see below), one still obtains
a double-well potential. However, as shown in Fig.3, the free
energy profile $\beta_0(\psi)$ is now non-degenerate.
Phase separation of the initial surface still occurs, but
the final phase separated stable orientations is now given
by the two points on the free energy profile sharing a common
supporting tangent (Fig.3).
Note that this dependence of the free energy profile on
initial surface orientation is
uncharacteristic of the usual spinodal decomposition where the
form of the coarse-grained free energy density
does not depend on the quench position.
Fig.2(b) illustrates the faceting patterns for
an off-critical quench with original surface tilted
$\theta_v=10^{\circ}$ from the $(11)$ direction, at $\epsilon=0.2$.

\FIG\three{Profile of the projected free energy density
$\beta_0(\psi)$ in \ftil\ for a crystal surface
tilted $\theta_v=10^{\circ}$ from the $(11)$-direction.
Anisotropy strength $\epsilon=0.2$.}

In our model, a surface sufficiently tilted from
the $(11)$-direction for fixed $\epsilon>\epsilon_c$ will
 be metastable rather than unstable. The
borderline tilting angle $\theta_{0}$ separating these two cases
is determined from the spinodal condition
$$\frac{\pa^2\beta_{0}}{\pa \psi^2}\bigg|_{\psi=0,\theta_v=
   \theta_{0}}= \frac{\pa^2}{\pa\psi^2}\left(
\frac{\gamma_{0}(\theta+\theta_v)}{\cos \theta}
\right)\bigg|_{\theta=0,\theta_v=\theta_{0}}=0. \eqn\spno $$
Eq.\spno\ is equivalent to the condition of zero surface
stiffness
$$  \gamma(\theta_{0})+\gamma''(\theta_{0})=0 \eqno\eq $$
or
$$ \theta_0=\frac{1}{4}\arccos \left(\frac{1}{15\epsilon}
\right). \eqno\eq $$
which gives for $\epsilon=0.2$,
$\theta_{0}=17.63^{\circ}$.
For tilting angles lying outside the spinodal region, i.e.,
$\theta_v >\theta_0$, one generally encounters a so-called
nucleation regime: the crystal surface will not phase separate
{\it spontaneously}
but will have to overcome a finite free energy barrier.
It must be emphasized that this nucleation regime has to be
treated as an intrinsically two-dimensional problem.
A nucleated facet is necessarily compact at the beginning.
Subsequent growth is sometimes seen to exhibit strong anisotropy,
giving rise to a long, slender facet. The speed of spreading
of such a facet along its axis was considered by Stewart and
Goldenfeld\rlap.\refmark{\stewart}
On the other hand, Mullins\refmark{\etch}
concentrated on the time-dependent widening of a linear facet
perpendicular to the facet axis. This
problem can also be studied straightforwardly using the equations
derived above.
We first create a large fluctuation,
mimicing a well-nucleated facet, on a metastable surface
($\theta_v=20^{\circ}>\theta_0$). Subsequent evolution of
the fluctuation is monitored by numerically simulating the
governing solution \criticsurf.
A time sequence of this development is presented in Fig.4.
We point out the particularly interesting
feature in Fig.4: as the facet widens, it induces at
its wake nucleations of additional facets from the metastable
surface.

\FIG\four{Propagation of a facet on a metastable crystal
surface, governed by \criticsurf.
The crystal surface is mis-oriented $20^{\circ}$
from the crystal $(11)$-direction. Time sequences
are $t=0$, $20$, $160$, and $640$. $\epsilon=0.2$.}

\chapter{TWO-DIMENSIONAL SURFACE ORDERING}
We now investigate the dynamics of a two-dimensional crystal
surface of fixed orientation.
Recalling that the order parameter is a 2-vector,
equation \eqmo\ can be written separately for the
two cases of distinct transport mechanisms.
We obtain
$$ \frac{\pa \bm\psi(\vecr,t)}{\pa t}
   =-A \nabla\left[ \nabla\cdot\left(a\nabla^2\bm\psi\rt -
      \frac{\pa \beta_{0}(\bm\psi,\bm\psi^*)}{\pa \bm\psi^*}
      \right)\right] \eqn\twodim $$
for evaporation-condensation dominated ordering, and
$$ \frac{\pa \bm\psi(\vecr,t)}{\pa t}
   =B\nabla^2\nabla\left[ \nabla\cdot\left(a\nabla^2\bm\psi\rt -
      \frac{\pa \beta_{0}(\bm\psi,\bm\psi^*)}{\pa \bm\psi^*}
      \right)\right] \eqn\twodims $$
for ordering through surface diffusion.
Note that since $\bm\psi$ is related to the surface height via
$\bm\psi=\nabla z\rt$, one must have $\nabla\times\bm\psi=0$ which is
automatically preserved by the above equations of motion.
Moreover, while \twodim\ describes non-conserved
dynamics, the additional Laplacian in \twodims\ makes the
order parameter in this case manifestly conserved.

It is again instructive to first examine the projected free energy
landscape $\beta_{0}$.  Consider a model free energy
 density appropriate for a solid with cubic symmetry,
$$ \gamma_{0}({\tilde\theta},{\tilde\phi})=\sigma_0\left[1+\epsilon
  \sin^2{\tilde\theta}
 (\cos^2{\tilde\theta}+\sin^4{\tilde\theta}\sin^2
  {\tilde\phi}\cos^2{\tilde\phi})\right] \eqno\eq $$
where the spherical angles ${\tilde\theta},{\tilde  \phi}$
are measured with respect to the
crystal symmetry axes, and $\epsilon >0$ is the anisotropy strength.
As before, different choice of initial surface orientation
results in different ordering dynamics. For simplicity,
we will only consider in this paper the dynamics of a
large surface whose normal is parallel to the crystal
$(111)$-direction\rlap.\Ref\shore{The ordering dynamics of a $(111)$
surface has also been studied recently by J.D. Shore, et al in a
discrete ``tiling model''.  J.D. Shore, M. Holzer and J.S.
 Sethna, \journal Phys. Rev.  B&46&11376(92).}
 This geometry is relevant to the experiments on the
coarsening of $(111)$ surfaces of NaCl crystals\rlap.\Ref\nacl{
D. Knoppik and A. Losch, \journal J. Cryst. Growth &34&332(76);
D. Knoppik and F.-P. Penningsfeld, \journal ibid. &37&69(77).}
Choosing $z$-axis along  the $(111)$-direction and $x$-axis along
the $(1{\bar 1}0)$-direction, we have for the projected
surface free energy density
 $$ \beta_{0}(\bm\psi,\bm\psi^*)=
  \sigma_0(1+|\bm\psi|^2)^{1/2}\left[ 1+\epsilon
  \sin^2{\tilde\theta}
 (\cos^2{\tilde\theta}+\sin^4{\tilde\theta}\sin^2
  {\tilde\phi}\cos^2{\tilde\phi}) \right]   \eqn\freenacl $$
where ${\tilde \theta},{\tilde\phi}$ are related to the
order parameter through
$$  {\tilde\theta}=\arccos\frac{1}{\sqrt{3}}\frac{1-\sqrt{2}q}{
    \sqrt{1+|\bm\psi|^2}} ; \,\,
{\tilde\phi}= \arctan\frac{\sqrt{2}+q+\sqrt{3}p }
 {\sqrt{2}+q-\sqrt{3}p}.  \eqno\eq $$
Proceeding as before by using the Frank diagram,
we can easily show that for small anisotropy
$0<\epsilon < 1.180$ the $(111)$-direction is contained in the
ECS and therefore our surface is stable. For larger anisotropy
 $\epsilon> 1.180$, the $(111)$-direction completely disappears
from the ECS resulting in a sharp corner in that direction.
But this statement alone does not establish definitively
the {\it local} stability of our surface. In fact, the surface
can be either metastable or unstable, depending further on the
strength of crystalline anisotropy. To see this, let us
perform a linear stability analysis of the surface. Consider a small
perturbation applied to the otherwise smooth crystal surface
$\delta\bm\psi\rt=\delta\bm\psi_{{\VEC k}}\exp
  (i{\VEC k}\cdot\vecr+\omega_{{\VEC k}}t)$. By linearizing
\twodim\ and \twodims\  using \freenacl, we obtain
 $$ \omega_{{\VEC k}}\delta\bm\psi_{{\VEC k}}
  =  (A+Bk^2)(b-ak^2) \left( \eqalign{
       &  k_x^2   \qquad \,\, k_xk_y \cr
       &  k_xk_y  \qquad \,\, k_y^2  } \right)
    \delta\bm\psi_{{\VEC k}} \eqno\eq $$
where $b=\frac{8\sigma_0}{27}(\epsilon-\frac{27}{16})$,
which readily gives the dispersion relation
 $$ \omega_{{\VEC k}}= (Ak^2+Bk^4)(b-ak^2).  \eqno\eq $$
Hence, for $b<0$, i.e., $ 1.180< \epsilon <\frac{27}{16}$,
our surface is metastable. Only when $\epsilon>\frac{27}{16}$,
does the surface become linearly unstable against
the spontaneous formation of hill-and-valley patterns.
This result, on the one hand, is interesting because it
demonstrates the close proximity of metastability and
absolute instability for a surface of {\it fixed} orientation.
On the other hand, it also means that it is difficult to
differentiate between these two regimes when
interpretating experimental observations.

Equations \twodim\ and \twodims\ can be regarded as special
forms of the continuum clock models\refmark{\clockmodel}
because $\beta_{0}(\bm\psi)$ usually contains multiply
degenerate minima.
These minima correspond to degenerate stable states with
values of the order parameter pointing at equivalent directions
of a clock.
In \freenacl\ with
$\epsilon>\frac{27}{16}$, $\beta_0$ contains three degenerate minima
corresponding to three stable surface orientations.
The degree of degeneracy is in general determined by
both the underlying crystalline symmetry and by
the initial orientation of the surface.  In our example,
a slight misorientation of the initial surface away from
the $(111)$-direction will lift the three-fold degeneracy
of the ground states.
It is precisely this strong dependence of $\beta_0(\bm\psi)$
on initial surface orientation, and the proximity of
metastability and instability even for the same orientation
as previously mentioned,
which afford extremely rich dynamics to \twodim\ and \twodims\
for 2D crystal surface.
It is not the subject of this paper to fully explore all
aspects of the complex dynamics. Rather we will concentrate
on the dynamics of unstable growth at long times, in particular,
on the asymptotic growth laws for the typical domain sizes $L(t)$
for an unstable surface.

It is by now well established that, for pure unstable systems
at least, the late stage ordering kinetics is controlled by
the interaction and annihilation of topological
defects\rlap.\Ref\mycon{Fong Liu and G.F. Mazenko,
\journal Phys. Rev. B&45&6989(92);
\journal Phys. Rev. B&46&5963(92).}
For the clock model, the lowest energy topological
excitations are vortices inter-connected by
interfaces\rlap.\refmark{\clockmodel,}\Ref\kawa{
K. Kawasaki, \journal Phys. Rev. A &31&3880(85).}
In the NaCl experiments\rlap,\refmark{\nacl}
the three-faced pyramids observed on
coarsening $(111)$ surfaces are vivid realizations of those
topological defects with corners mapping to vortices and
edges to interfaces, respectively.  Reaffirming this analogy
between 2D crystal problem and the dynamics of clock model,
we anticipate that interactions between corners and edges
should dominate long-time asymptotic dynamics
of the crystal surface.

\FIG\five{Hill-and-valley patterns during the evolution of
a 2D unstable surface $z(\vecr,t)$, obtained
from numerical simulations
of model \twodim\ on $64\times 64$ lattices. Figures from
(a) to (d) are sequential in time at $t=50$, $200$, $800$, and
$3200$.  Vertical axis is in arbitrary scale.}

We have carried out preliminary numerical simulations of
the dynamics \twodim\ and \twodims, using explicit finite
difference scheme.  Again, by rescaling units of time and
length, we set the constants $A, B, a$, and $\sigma_0$ to
unity and obtain dimensionless equations of motion.
For most of the results reported
here, discretizations were performed on
square grids with sizes ranging from $64\times 64$ to
$256\times 256$ with grid spacing $\Delta x=1.5$.
Time increments were typically selected to be $\Delta t=0.1$.
Convergence of algorithm was checked by experimenting with different
values of $\Delta x, \Delta t$. It should be warned that the values
$\Delta x=1.5, \Delta t=0.1$ are in general too large to
probe accurately structures on the scale of a few
grid spacings. However, they were
used mainly to access efficiently the long-time growth regime.
Fig.5 depicts  typical
hill-and-valley structures of an evolving unstable surface,
taken from numerical solutions of \twodim\ on $64\times 64$ grids.

To extract the growth law, we show in Fig.6(a)
the typical pattern size $L(t)$ as a function of time on
a log-log plot.
Here the characteristic length scale $L(t)$ is defined as the
position of the first zero of the real space correlation function
$\VEV{\bm\psi\rt \cdot \bm\psi(0,t)}$. The average is over the
whole lattice with many independent realizations of initial
conditions.  An alternative definition of
$L(t)$, as the mean distance between corners,
has also be considered. Both lengths are shown to be
proportional to each other and therefore are equivalent.
Fig.6(a) corresponds to evaporation-condensation dynamics \twodim\
with surface free energy density \freenacl\ under anisotropy strength
$\epsilon=3.0$, simulated on $256\times 256$ lattices and averaged
over $20$ independent initial conditions.
It can be seen from Fig.6(a) that $L(t)$ is well
fitted by a power law $L(t)\sim t^n$,
with the exponent given by a linear regression estimate of
$n=0.23\pm 0.01$.
Simulations are also done for the ordering dynamics \twodims\
of surface-diffusion mechanism. The result is displayed in
Fig.6(b), again with anisotropy strength $\epsilon=3.0$.
In this situation the dynamics are much slower than in the previous
case and a longer transient is observed.
However, $L(t)$ can still be reasonably described by
the power-law form. The linear regression analysis gives
in this case $n=0.13\pm 0.01$. This value seems to be roughly
consistent with the result of Shore \etal\refmark{\shore} in their
simulations of a discrete tiling model.

\FIG\six{Characteristic domain size $L(t)$ vs $t$ on
  a log-log plot. Results are from simulations on $256\times
256$ lattices averaged over $20$ random initial conditions.
  (a) Dynamics from model \twodim, evaporation-condensation
  mechanism; (b) Dynamics from model \twodims, surface-diffusion
  mechanism.}

The values of growth exponent we obtained $n=0.23, 0.13$
for both transport mechanisms are smaller than the estimates
$n=1/2, 1/4$ reached by Mullins-type arguments\rlap.\refmark{
\etch}
This result reflects the highly nonlinear nature of
the equation of motions. Furthermore, interestingly, both
values can be roughly understood from a naive dimensional analysis
of the equations of motion \twodim\ and \twodims. Direct power
counting of both sides of the equations gives the estimate
$n=1/4$ and $n=1/6$ for the two transport mechanisms respectively,
which are rather close to the numerical values we determined.
In growth kinetics, different kinds of defect structures
affect in varying degrees the asymptotic dynamics. The
curvature-driven motion of interfaces usually provides more
efficient ordering than the motion of vortices does.
We believe that the agreement on $n$ between our simulations
and naive dimensional analysis
is indicative that both corners and edges are playing
active roles in the coarsening dynamics within the
time regime we have studied.
We must point out, however, that this picture can change.
At longer times the system may crossover to another growth
regime dominated by the curvature-driven motion of edges alone,
with distinct growth exponents.
Our simulations were not efficient enough to probe into
this possibility, which is deferred to future studies.

Extended power-law growth of the characteristic length scale
hints at the possibility of dynamical scaling. Here we
concentrate on the order-parameter correlation function
$\VEV{\bm\psi\rt \cdot \bm\psi(\vecr',t)}$. Dynamical
scaling ansatz states that at long times, the order-parameter
correlation function satisfies the self-similarity relation
  $$  \VEV{\bm\psi\rt \cdot \bm\psi(\vecr',t)}=
  \psi_0^2 F\left(\frac{|\vecr-
     \vecr'|}{L(t)}   \right) \eqn\ansatz  $$
for a homogeneous system, where $\psi_0$ is the equilibrium
magnitude of the order parameter
and  $F$ is a {\it universal} function.
Scaling hypothesis \ansatz\ is indeed borne out reasonably
 well by our numerical simulations.
Order-parameter correlation functions at different times were
evaluated and plotted according to the ansatz \ansatz\ in
Fig.7 for the two dynamics \twodim\ and
\twodims. In each case, data are collected
from $20$ simulations on $256\times 256$ lattices.

Geometries and correlations of crystal surfaces can be
probed by diffraction measurements or by direct-imaging
techniques, the result of which can be directly compared
with that from theoretical calculations. Here we merely note
that one of the experimentally measurable quantities, the
height-height correlation function, is simply related to the
order parameter correlation function through
  $$ \VEV{\bm\psi\rt \cdot \bm\psi(\vecr',t)}=- \nabla^2
    \VEV{z\rt z(\vecr',t)}  \eqno\eq $$
or, in terms of the Fourier representation,
  $$ \VEV{\bm\psi_{\VEC k}(t)\cdot\bm\psi_{-\VEC k}(t)}=k^2
  \VEV{z_{\VEC k}(t)z_{-\VEC k}(t)}.  \eqno\eq $$
These correlation functions have interesting short-distance
properties related to local structures of the topological
defects\rlap.\refmark{\mycon}

\FIG\seven{Demonstration of dynamical scaling of the order-parameter
correlation function. (a) and (b) correspond to models \twodim\
 and \twodims, respectively.}

\chapter{DISCUSSIONS}

The dynamical evolution of an unstable crystal surface has
been studied using a continuum approach. The principle
objective of this paper has been to demonstrate the importance
of nonlinearity and crystalline anisotropy in unstable surface
ordering, and to illustrate a close relationship  between
the problem of surface dynamics and that of phase-ordering
dynamics in bulk systems.
We have shown that certain concepts in phase transition kinetics
are equally useful here in describing phenomena of crystal surface
kinetics: it is revealing to regard stable surface
orientations as genuine coexisting thermodynamic phases, and
thereby to realize the crucial role of phase boundaries or
topological defects in controlling the asymptotic ordering
dynamics. Our continuum method, focusing on larger time and space
scales, is complementary to the more common microscopic approaches
on crystal growth kinetics\rlap.\Ref\atomistic{For references,
see J.D. Weeks and G.H. Gilmer, \journal Adv. Chem. Phys. &40&157(79);
A. Madhukar and S.V. Ghaisas, \journal CRC Crit. Rev. Solid State
Mater. Sci. &13&1434(87); D.D. Vvedensky, \etal, in
{\sl Kinetics of Ordering and Growth at Surfaces}, edited by
M.G. Lagally (Plenum, New York, 1990); H. Metiu, Y.T. Lu and
Z. Zhang, \journal Science &255&1088(92).}

Certain simplifications and approximations have been made
in this work, which will be further investigated in the future.
First and foremost, we have, for simplicity,
tentatively neglected any
elastic effects in our treatment. In reality, the presence of
surface stresses induces compensating
volume stresses in the crystal which must be
included in any realistic calculations.
There exist evidences that elastic effects couple
to surface structural properties. The importance of
stress relaxation in surface ordering dynamics has been
addressed theoretically by a number of
authors\rlap.\refmark{\stewart,}\Ref\marchen{V.I. Marchenko,
\journal Sov. Phys. JETP &54&605(81); O.L. Alerhand, \etal,
\journal Phys. Rev. Lett. &61&1973(88); O.L. Alerhand,
\etal, \journal Phys. Rev. Lett. &64&2406(90).}
In our problem, the
energy density (per unit area) decrease due to the elastic effect
can be shown to scale roughly as\refmark{\marchen}
 $$E_e \sim  -\frac{\gamma_e \xi_0}{L} \ln \left(\frac{L}{
    \xi_0}\right)   \eqno\eq $$
where $\gamma_e$ is some typical elastic energy scale
and $\xi_0$ is the edge thickness. This quantity, combined with the
excess surface free energy due to the presence of edges which
scales as $E_s \sim \gamma_0\xi_0/L$, gives the total
energy density:
  $$ E_{total} \sim \frac{\gamma_0\xi_0}{L}-
            \frac{\gamma_e\xi_0}{L}\ln \left(\frac{L}{\xi_0}
         \right).  \eqno\eq $$
Now it immediately follows that the elastic effect  will eventually
stablize the coarsening at a length scale $\xi_e \sim
\xi_0 \exp (1+\gamma_e/\gamma_0)$. The results
in this paper will therefore be valid, if we presuppose
that $\xi_e$ is much larger than the length scales accessed in
our present study. In passing, we want to address a common
misconception regarding the importance of elstic effects in
surface ordering.  It is sometimes
asserted\rlap,\Ref\herringelast{For one example, see discussions
by Herring in Ref.\herring.} by comparing
the elastic relaxation energy $E_e$ with the surface free energy
density $\gamma_0$ in terms of the
ratio $E_s/\gamma_0 \sim L^{-1}\ln L$, that the elastic
contribution scales inversely with the system size therefore is
negligible.  This argument is flawed when applied to our
unstable surface ordering, because the relevant energy scale
 for $E_s$ to
compare to is {\it not} $\gamma_0$ but rather the excess
free energy $E_s$ associated with topological defects.

In deriving the equation of motion, the kinetic
transport coefficients were assumed to be isotropic.
A more realistic choice of kinetic coefficients, as well as
other factors neglected in the present analysis such as
thermal fluctuations, external deposition flux, impurity
adsorption, alternative transport mechanisms,
can all be straightforwardly incorporated into the model.
The effect of some of these factors can introduce additional
features to the dynamics of surface evolution.

Finally, attention must be paid to the form of
surface free energy density used.
In the forgoing sections, we have
chosen $\gamma({\bm n})$ as being
smooth functions of the surface orientation. Strictly
speaking, one must also consider the possibility that
the polar plot of $\gamma$ in general may
contain cusps in directions corresponding to surfaces
with small Miller indices,
even at temperatures approaching the bulk melting temperature.
The presence of these cusps introduces extra stability to
some facet surfaces and therefore complicates the dynamics.
However, this does not post conceptual problems. A simple
regularization procedure, and certain modification of the
equations of motion may well suffice to circumvent
the singularity near those cusps. This question will also
be addressed in the future.

\bigskip
\ack
We are grateful to Prof. J.S. Langer and
Prof. N. Goldenfeld for encouragements and for helpful
discussions. F.L. thanks J. Stewart, Dr. A.C. Shi and
Dr. J. Shore for discussions and for providing useful
references.  This work was supported in part by the NSF Grant
PHY89-04035, and the NSF Science and Technology Center for
Quantized Electronic Structures (Grant DMR91-20007).
\endpage
\refout\endpage
\figout
\end